\documentclass{article}
\usepackage{authblk}
\usepackage[a4paper, margin=1in]{geometry}
\usepackage[utf8]{inputenc}
\usepackage[T1]{fontenc}
\usepackage[numbers]{natbib}
\usepackage{amsmath}
\usepackage{graphicx,todonotes}
\usepackage{multirow}
\usepackage{algorithm}
\usepackage{algpseudocode}
\usepackage{listings}
\usepackage{xcolor}

\colorlet{punct}{red!60!black}
\definecolor{background}{HTML}{EEEEEE}
\definecolor{delim}{RGB}{20,105,176}
\colorlet{numb}{magenta!60!black}

\lstdefinelanguage{json}{
    basicstyle=\normalfont\ttfamily,
    showstringspaces=false,
    breaklines=true,
    frame=lines,
    backgroundcolor=\color{background},
    literate=
     *{0}{{{\color{numb}0}}}{1}
      {1}{{{\color{numb}1}}}{1}
      {2}{{{\color{numb}2}}}{1}
      {3}{{{\color{numb}3}}}{1}
      {4}{{{\color{numb}4}}}{1}
      {5}{{{\color{numb}5}}}{1}
      {6}{{{\color{numb}6}}}{1}
      {7}{{{\color{numb}7}}}{1}
      {8}{{{\color{numb}8}}}{1}
      {9}{{{\color{numb}9}}}{1}
      {:}{{{\color{punct}{:}}}}{1}
      {,}{{{\color{punct}{,}}}}{1}
      {\{}{{{\color{delim}{\{}}}}{1}
      {\}}{{{\color{delim}{\}}}}}{1}
      {[}{{{\color{delim}{[}}}}{1}
      {]}{{{\color{delim}{]}}}}{1},
}
\definecolor{blueish}{HTML}{0089cc}
\definecolor{blueish_dark}{HTML}{005076}
\usepackage{subcaption}
\usepackage[hyphens]{url}

\title{AiCEF: An AI-assisted Cyber Exercise Content Generation Framework Using Named Entity Recognition}
\begin{document}
\author[1]{Alexandros Zacharis}
\author[2,3]{Constantinos Patsakis}

\affil[1]{European Union Agency for Cybersecurity (ENISA)}
\affil[2]{Department of Informatics, University of Piraeus, 80 Karaoli \& Dimitriou str., 18534 Piraeus, Greece}

\affil[3]{Information Management Systems Institute of Athena Research Centre, Greece}

\date{}
\maketitle

\begin{abstract}
Content generation that is both relevant and up to date with the current threats of the target audience is a critical element in the success of any Cyber Security Exercise (CSE). Through this work, we explore the results of applying machine learning techniques to unstructured information sources to generate structured CSE content. The corpus of our work is a large dataset of publicly available cyber security articles that have been used to predict future threats and to form the skeleton for new exercise scenarios. Machine learning techniques, like named entity recognition (NER) and topic extraction, have been utilised to structure the information based on a novel ontology we developed, named Cyber Exercise Scenario Ontology (CESO). Moreover, we used clustering with outliers to classify the generated extracted data into objects of our ontology. Graph comparison methodologies were used to match generated scenario fragments to known threat actors' tactics and help enrich the proposed scenario accordingly with the help of synthetic text generators. CESO has also been chosen as the prominent way to express both fragments and the final proposed scenario content by our AI-assisted Cyber Exercise Framework (AiCEF). Our methodology was put to test by providing a set of generated scenarios for evaluation to a group of experts to be used as part of a real-world awareness tabletop exercise.
\end{abstract}

\section{Introduction}
Cyber Security Exercises (CSE) are increasingly becoming an integral part of the cybersecurity training landscape \cite{karjalainen2019pedagogical} providing a hands-on experience to personnel of both public and private organisations worldwide.
A CSE, as described in the ISO Guidelines for Exercises\cite{ISO22398:2013}, is "\textit{a process to train for, assess, practice, and improve performance in an organisation}". ENISA defines a CSE as "\textit{a planned event during which an organisation simulates cyber-attacks or information security incidents or other types of disruptions to test the organisation's cyber capabilities, from being able to detect a security incident to the ability to respond appropriately and minimise any related impact.}" \cite{conklin2006cyber}.
\subsection{Problem setting and objectives}

The creation of CSE content is a painstaking process that requires a deep understanding of the current threat landscape and the historical threats and incidents faced by an entity and the corresponding sector. Furthermore, training employees with simulated incidents is the closest method to testing the preparedness and effectiveness of measures and procedures set in place. Creating relevant and dynamic content for developing CSE scenarios requires expertise and resources often lacking among most organisations.

The main objective of our work is automating generation of structured CSE scenarios based on a pool of unstructured information with little experience in scenario building expected from the Exercise Planner (EP).

The standard method for preparing an exercise scenario \cite{ISO22398:2013} lays down three layers, namely events, incidents, and injects. After developing a scenario, an organisation must ensure that it contains only necessary information. Moreover, it must be designed to test participants' capabilities in a stressful environment. Events, at the first level, provide the general description of an exercise scenario. Depending on previously decided objectives and aims, the number of events can differ from one exercise to another. Each event would have a specific set of consequences at the second level. These consequences are called incidents. An event can have multiple consequences, which can affect each other. On the third level, injects facilitate the communication of events and incidents to the exercise participants. An ideal inject would provide exercise information and problems to be solved. At the same time, it would indirectly force participants to act on those consequences and make decisions.

The proposed scenarios should satisfy the specifications provided by the EP. Such specifications can be the training topics and objectives, the sector to focus on or specific threats of interest that are currently or will be trending in the future. For simplicity, in what follows, when referring to sectors, we will refer to the ones of NIS2 \cite{nis2}; however, any other such classification can be used.
More specifically, the objectives can be summarised as follows:
\begin{enumerate}
\item Create an ML-powered Exercise Generation Framework that would:
    \begin{enumerate}
        \item Generate structured exercise scenarios that reflect an organisation's current or future threat landscape, including potential threat actors and the corresponding techniques, tactics, and procedures (TTPs).
        \item Generate scripted events and incidents that could materialise in the context of a real attack against an organisation (belonging to any NIS2 defined Sector)
        \item Identify and describe artefacts that could accompany the exercise scenarios as potential injects
    \end{enumerate}
\item The generated scenarios should be expressed in a structured way or format, following an Ontology. The generated outputs should be both machine and human-readable.
\item The proposed methodology and tools created should provide qualitative and quantitative added value in CSE development and cyber-awareness by measuring the following Key Performance Indicators (KPIs):
    \begin{enumerate}
        \item Improve the speed in CSE generation (quantitative)
        \item Improve quality in CSE generation (qualitative) for inexperienced EPs
        \item Improve the relevance of proposed CSE scenarios to the current threat landscape (qualitative)
    \end{enumerate}
\end{enumerate}

The use of case studies will help measure the results of the KPIs set by comparing the traditional exercise generation methods and tools versus the proposed ones through an evaluation provided by an Ad-hoc Cyber Awareness Expert Group \footnote{"https://www.enisa.europa.eu/topics/cybersecurity-education/ad-hoc-working-group-awareness-raising"} \footnote{"The information and views set out in this report are those of the author(s) and do not necessarily reflect the official opinion of the European Union Agency for Cybersecurity (ENISA). Neither the European Union institutions nor any person acting on their behalf may be held responsible for any use that may be made of the information contained therein} that will peer review the outputs of the aforementioned methodology.

\subsection{Main contributions}
The contribution of this work is twofold. Initially, we predict future cyber-attack trends and the overall threat landscape against specific organisation sectors and propose customised awareness training topics by clustering them into training topics. Then, we automate the process of generating the corresponding content for cyber awareness exercises with machine learning (ML).

\begin{figure}[th]
    \centering
    \includegraphics[width=.7\linewidth]{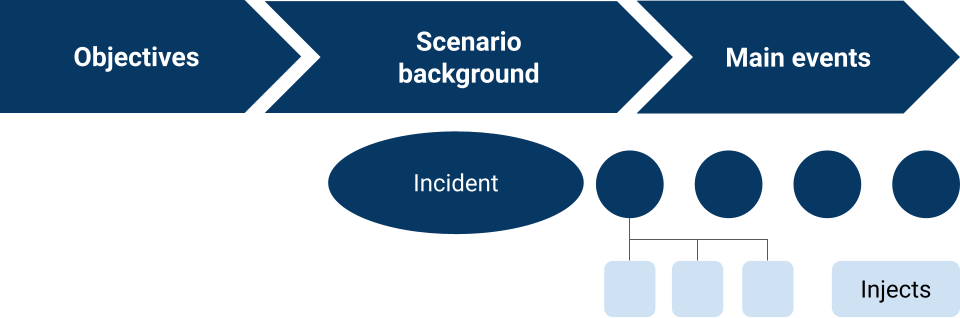}
    \caption{Cyber Exercise Structure as per ISO 22398:2013 \cite{ISO22398:2013}.}
    \label{fig:iso_22398}
\end{figure}

Our proposed methodology, which a set of tools will accompany, allows an inexperienced EP to fully structure CSE scenarios from free text following our proposed Cyber Exercise Scenario Ontology (CESO). The exercise structure will follow the traditional Scenario- Events-Incidents-Injects tree structure ISO 22398:2013 \cite{ISO22398:2013} as depicted in Figure \ref{fig:iso_22398}. Additional cyber exercise content will be generated to complement the scenario and proposals for the fittest of a set of given training topics to better prepare an organisation for an imminent cyber crisis.

Through our work, we fill the gap in the lack of expertise by the average cyber security expert that acts as an Exercise Planner and provide the tools and the methodology to design CSE scenarios in an easy, automated, and structured way. To achieve this we combine the power of Machine Learning (ML) and more specifically, Named Entity Recognition (NER), with a set of novel Cyber Exercise Scenario Ontology (CESO) and CSE scenario generation framework dubbed AiCEF. Finally, an evaluation Methodology and its results are presented in along with ideas for future work.

\section{Related work}
CSEs, also known as Cyber Defense Exercises (CDX), have been considered as a effective way to implement an engaging security awareness training \cite{furtunua2010structured,schweitzer2009active} experience. CSEs have been characterised as a highly effective method to provide an ultimate learning experience \cite{augustine2006cyber}, helping individuals or teams of varying expertise, improve a range of skill related to informational security. Furthermore, via exercising, organisations can uncover gaps in security policies, procedures and resources  \cite{dewar2018cybersecurity,gurnani2014scalable} leading to awareness training, tools and policies improvements.

Previous work in the CSE domain \cite{schepens2002cyber} has highlighted the use of cyber defence competitions or live-attack exercises as a very effective way of teaching information security \cite{dodge2004organized,karagiannis2021engaging}, helping teams design, implement, manage and defend a network of computers \cite{adams2009collective,conklin2005use,conklin2006cyber,mullins2007cyber,mullins2007impact}.
Vigna \cite{vigna2003teaching} and Mink \cite{mink2006attack} further support these findings.

Further research was conducted on cyber defence competitions \cite{rursch2009adventures,white2010cyberpatriot} and the most suited architecture \cite{schepens2003architecture} and tools and techniques to be used in order to create an active learning experience were described by Green et al. \cite{green2013addressing}. Patriciu and Furtuna \cite{patriciu2009guide} presented several steps and guidelines to be followed when designing a CSE. White \cite{white2004cyber} introduced a different approach to such live CSEs, presenting lessons learned and providing suggestions to help organisations run their own exercises. Other works in the literature examined how to run CSEs, using a service provider model \cite{mattson2007cyber}.

CSEs can be used as a tool to generate scientifically valuable datasets for future security research \cite{sangster2009toward,sommestad2012cyber} and help uncover hidden risk from weak Security policies and/or procedures \cite{samejima2012risk}. CSEs can even be used to measure performance against specific standards \cite{dodge2009standards} or team effectiveness based on behavioural assessment techniques \cite{granaasen2016measuring}.
Moreover experiments using various platforms like the RINSE simulator \cite{liljenstam2006rinse} or using realistic inter-domain routing experiment platform \cite{li2009real}, for rendering of network behaviour.

Focusing further on the human aspect, Job Performance Modelling (JPM) using vignettes for improving cybersecurity talent management through cyber defence competition design, was described by Tobey \cite{tobey2015vignette}.

A successful CSE counts heavily on the use of a robust scenario.
Exercise scenarios must describe worst-case scenarios that participants can relate to and are realistic enough to trigger seamless engagement.
Intuitive scenarios can be a powerful tool that can predict future states or situations \cite{augustine2006cyber},\cite{furtunua2010structured}; incorporating issues to be resolved, interactions and consequences \cite{green2013addressing}, \cite{granaasen2016measuring} leading to a constructive training experience.

An exercise's scenario is a sequential, narrative account of a hypothetical incident that provides the catalyst for the exercise and is intended to introduce situations that will inspire responses and thus allow demonstration of the exercise objectives \cite{schepens2003architecture}.
In the context of CSEs, a scenario defines the training environment that will lead participants towards fulfilling the exercise objectives \cite{kick2014cyber} set.
The cyber security problem described in a scenario itself portrays a structured representation, named Master Scenario Events List (MSEL), which serves as the script for the execution of an exercise \cite{schepens2003architecture}.
CSE scenarios formats can vary \cite{planning2008directors} but two are the most prevalent:
\begin{itemize}
\item Outlined scenarios: Provide a general summary of the impact of an event to assets. \cite{scarfone2008sp}
\item Detailed scenarios: Contain exhaustive information sequentially describing the event's impact on specific services or sections of an organisation, along with a timeline for restoring key functions. \cite{Cyber_Tabletop}
\end{itemize}

Recent trends in attack recognition utilise AI, ML, and NLP tools and techniques to empower their efficiency. However, there needs to be more dedicated methodology focusing on CSE scenario generation. There is a need for a methodologically built and annotated CE corpus that could train multiple algorithms for Cyber Exercise elements. Such a corpus should focus on the syntactic and semantic characteristics of the cyber exercise components and broaden our understanding of the malicious patterns used in cyber incidents that can be reused for CSE material. A similar approach to the one used in building and evaluating an annotated Corpus for automated Recognition attacks has been utilized \cite{tsinganos2021building}, only this time to extract CSE relevant objects.

Following Cyber Security related ontology creation examples \cite{pastuszuk2021cybersecurity}, ontology~-based scenario modelling for CSEs have already been proposed \cite{wen2021ontology}. Still, an ontology that is truly compatible with Machine Learning algorithms is missing and will be the focus of our work.

\section{Cyber Exercise Scenario Ontology (CESO)}
Our work so far highlighted the need for a common CSE scenario ontology for translating the various parts of an exercise while keeping a close link to popular already used ontologies for cyber incident representations. The analysis of the domain revealed many taxonomies for different areas of the cybersecurity domain (types of attacks, vulnerabilities, sectors, harm) but those needed to be linked together in a model that allows for an EP to represent a CSE accurately.

To build our ontology, the following questions were raised:
\begin{enumerate}
\item What is the scope of the ontology?
\item Should we consider reusing existing ontologies or taxonomies?
\item What are the important terms in the ontology?
\end{enumerate}

The scope of the ontology was determined by asking competency questions to experienced EPs that helped us identify the most important terms. We also used the domain expert’s knowledge to identify prominent existing ontologies and ways to reuse them.
The steps followed were:
\begin{enumerate}
    \item Define the scope of our Ontology
    \item Identify other ontologies or taxonomies that can be used/reused
    \item Define the main concepts and the relationships between them
    \item Define the properties of the concepts
    \item Implement the ontology
\end{enumerate}

\subsection{Scope}
The scope of the defined model was to target an efficient and robust way of representing cyber incidents in the context of a CSE. After all, a CSE is a collection of simulated incidents provided to players in an orchestrated way to achieve the exercise's objectives.

The exercise ontology presented is incident-centric, focusing on using a bottom-up approach that allows us to identify and describe incidents first so we can group them into Events and then cover the full generation of CSE scenarios that fit the high-level objectives set.

The first building blocks, incidents, are assigned injects and mitigation actions that match the expected scope of the scenario. Injection timing is configured on the attribute level of each object. As we build toward the higher level of the exercise the scenario is formed. The selected format should allow for the scenario's portability to various existing tools (ex. MISP\footnote{https://www.misp-project.org/}) and support a decentralised type of CSE execution.

\subsection{Ontologies/taxonomies to be (re)used}
A set of existing Ontologies, Taxonomies, Frameworks, Standards and Formats have been explored with relevance to Cyber Security and a focus on the representations of the key element of CSEs from the point of their very building blocks being the incidents to be simulated.
Our research concluded that a combination of the following would provide the necessary means: ISO 22398\cite{ISO22398:2013}, MITRE ATT\&CK\cite{MITRE} and Cyber Kill Chain\cite{cyberkillchain}, MITRE CVE\cite{mitrecve}, STIX 2.1\cite{stix}.

We chose Stix 2.1 as the basis for our ontology, which defines a taxonomy of cyber threat intelligence to be extended to cover our need to describe a CSE scenarios
The STIX2.1 model describes an adversary and adversary activities in appropriate data structures by default. STIX Domain Objects cover: Threat Actor; Malware; Tools; Campaign; Intrusion Set and Attack Pattern (referencing the Common Attack Pattern Enumeration and Classification taxonomy, CAPEC), perfectly covering what is called incident \& injects in the CSE nomenclature.
Moreover, STIX 2.1 enables organisations to share CTI in a consistent and machine-readable manner, allowing security communities to understand better what computer-based attacks they are most likely to face and anticipate and/or respond to those attacks faster and more effectively.

This helps us build on top of these communities to reuse existing tools and share CSE scenarios represented in the very same format.

\subsection{Scenario Augmented Model}
Based on the bottom-up approach, a Scenario Augmented Model (SAM) is proposed in two layers that cover both the Informational and Operational aspects with the same objects but utilise different attributes.

\begin{figure*}[th]
    \centering
    \includegraphics[width=.8\textwidth]{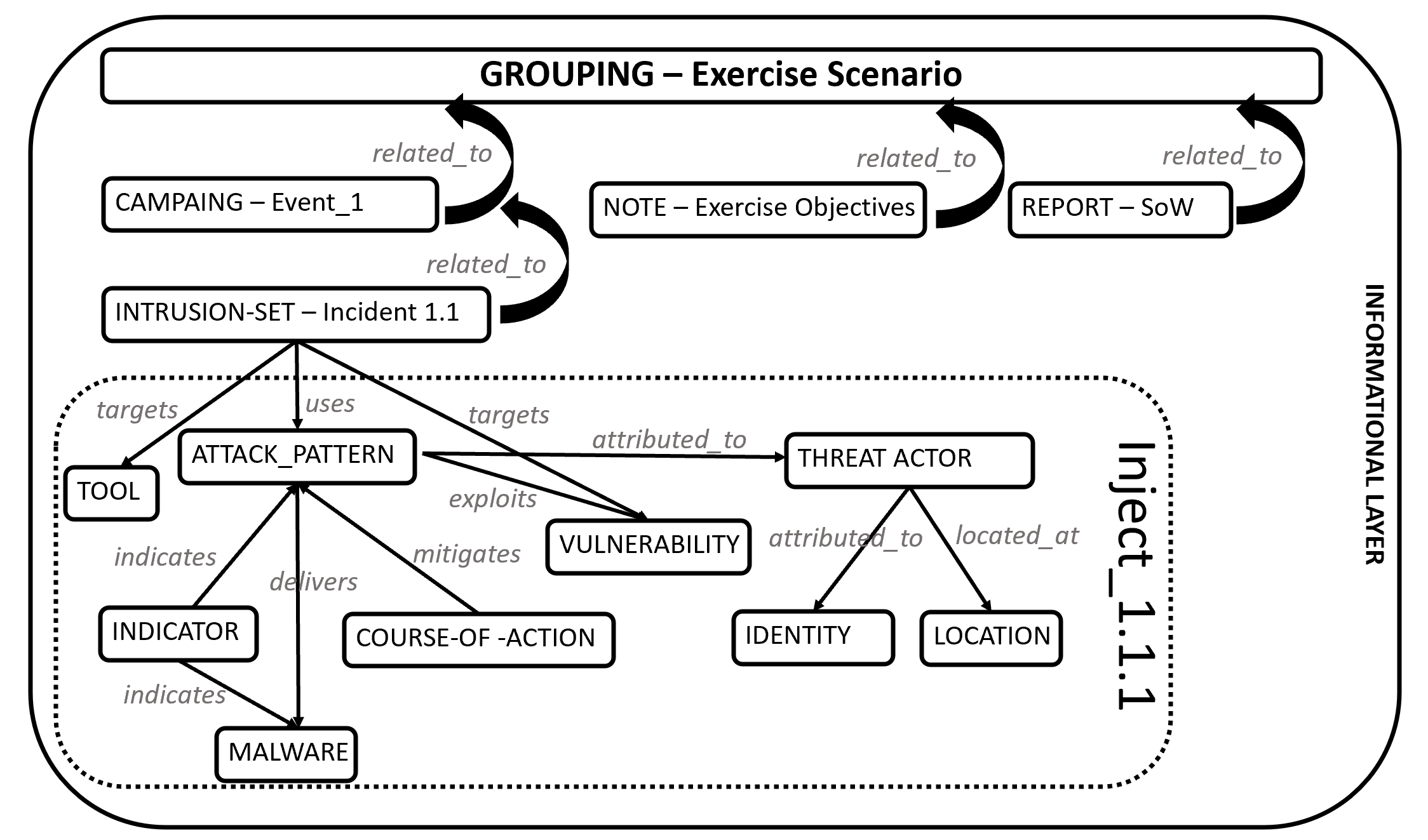}
    \caption{Informational Layer of CESO.}
    \label{fig:ceso_informational}
\end{figure*}

The Informational Layer covers the context and main attributes of scenarios. Figure \ref{fig:ceso_informational} describes the key relationships in the informational layer.

The whole exercise is grouped using the Grouping Objects. The object holds information related to the exercise's name, description and scenario. All Events, Objectives and State of the World (SoW) and their matching objects (Campaign, Note, Report) are related to the Exercise Scenario.

One or more Incidents (Intrusion Set) can be related to Events. From there, various objects with interlinked dependencies form the Inject in a Course of Actions Instance that refers to all related objects of an Attack Pattern.

An Inject can contain the following objects: Attack Pattern, Tool, Vulnerability, Indicator, Malware Threat Actor (who is attributed and Identity and is located at a Location) and a Course Of Action.
Injects do not have to be related to an Event or Incident. Such examples are the STARTEX or ENDEX \footnote{The [START]ing and [END]ing [EX]ercise injects}, which can be represented only with a Course of Action object but are directly related to the Scenario.

The Scenario Operational Layer describes an exercise scenario's execution flow, mainly dealing with injects delivery to the intended recipients. There are two major interrelated parts: (1) the events/injects, which describe the detailed activities of the scenario and expected actions from the participants, and (2) the Participants.

\begin{figure*}[th]
    \centering
    \includegraphics[width=.8\textwidth]{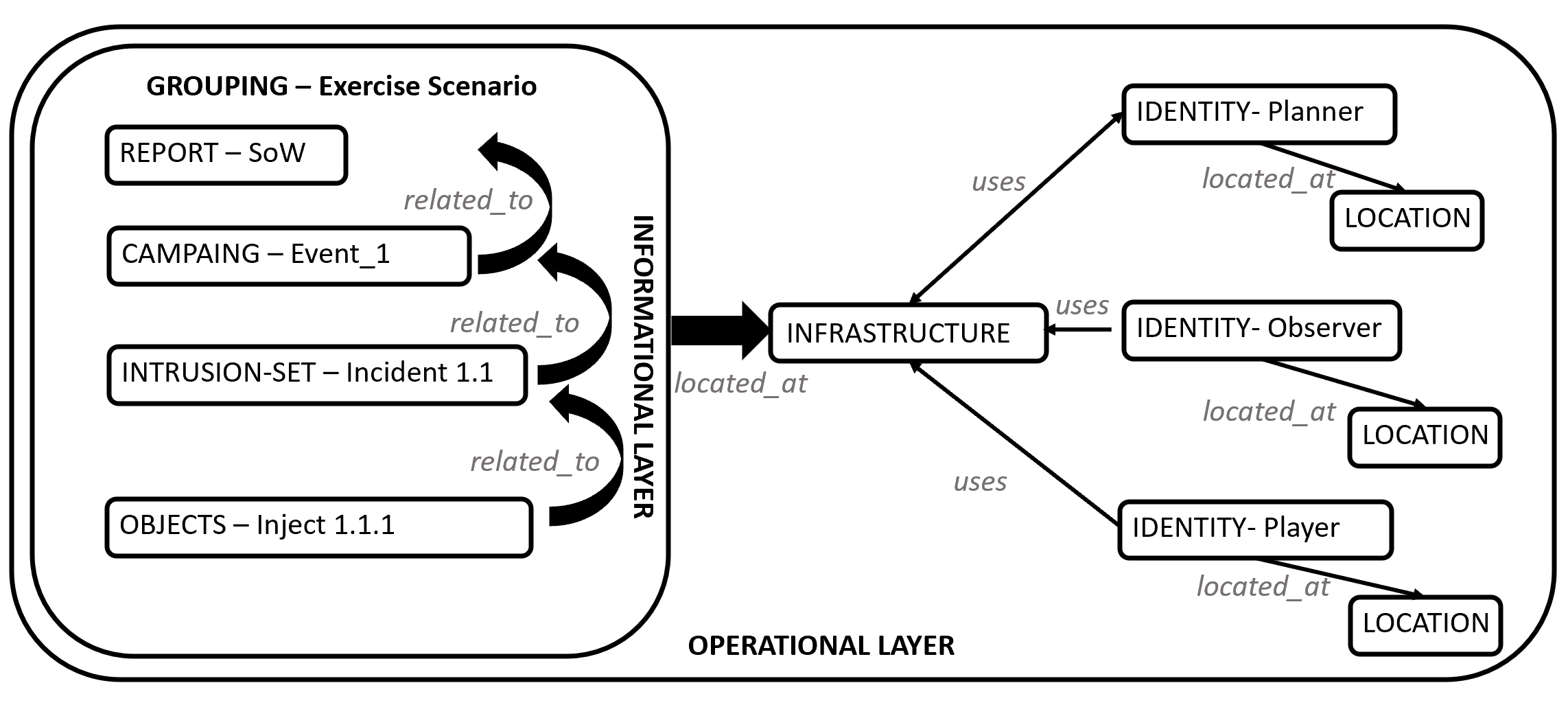}
    \caption{Operational Layer of CESO}
    \label{fig:ceso_operational}
\end{figure*}

The whole scenario, including Events, Incidents and Injects is stored in an Infrastructure object, representing the Exercise Platform. This platform is used by EPs (Identity) to design and conduct the exercise, Observers, and Players to interact with the Scenario. All Participants are located in the same or different Locations. The Operational Layer is illustrated in Figure \ref{fig:ceso_operational}.

\subsection{Implementing the Ontology}
Keeping the structure of CSE intact, the following STIX 2.1 Objects have been repurposed to fulfil our goal to represent the main CSE components successfully covering SAM along with matching relationships.

\textbf{Objects:} All STIX 2.1 defines object as per specifications.
\begin{table}[th]
    \centering
    \begin{tabular}{|p{2.5in}|p{3.5in}|}
    \hline
         \textbf{CSE Component}& \textbf{STIX 2.1 Object} \\
         \hline
         Exercise Details \& Scenario Background & Grouping\\
         \hline
         Objectives & Note\\
         \hline
         Events & Campaign\\
         \hline
        Incidents & Intrusion Set \\
        \hline
        State of the World (SoW) & Report\\
        \hline
        Injects & Tool, Vulnerability, Threat-Actor, Identity, Location, Attack Pattern, Malware, Indicator, Course-of-action, Observed Data, Malware Analysis, Report\\
        \hline
        Exercise Platform & Infrastructure\\
        \hline
        Exercise Participant & Identity, Location\\
        \hline
    \end{tabular}
    \caption{CSE Components to STIX 2.1 objects mapping}
    \label{tab:objects}
\end{table}

\textbf{Relationships:} All relationships are implemented as per STIX 2.1 relationship object specifications. The relationships in Table \ref{tbl:relationships} (representing the edges of the graph) have been identified between key objects, but more can be used.

\begin{table}[th]
\centering
\begin{tabular}{|l|l|l|}
\hline
\textbf{Source Object} & \textbf{Destination Object} & \textbf{Relationship} \\ \hline
Campaign & Grouping & \multirow{5}{*}{\textbf{related\_to}} \\ \cline{1-2}
Note & Grouping &  \\ \cline{1-2}
Report & Grouping &  \\ \cline{1-2}
Intrusion-Set & Campaign &  \\ \cline{1-2}
Course-Of-Action & Grouping &  \\ \hline
Intrusion-Set & Tool & \multirow{2}{*}{\textbf{targets}} \\ \cline{1-2}
Intrusion-Set & Vulnerability &  \\ \hline
Intrusion-Set & Attack-Pattern & \multirow{2}{*}{\textbf{uses}} \\ \cline{1-2}
Identity & Infrastructure &  \\ \hline
Attack-Pattern & Threat-Actor & \multirow{2}{*}{\textbf{attributed\_to}} \\ \cline{1-2}
Threat-Actor & Identity &  \\ \hline
Identity & Location & \textbf{located\_at} \\ \hline
Attack-Pattern & Malware & \textbf{delivers} \\ \hline
Attack-Pattern & Indicator & \multirow{2}{*}{\textbf{indicates}} \\ \cline{1-2}
Indicator & Malware &  \\ \hline
Attack-Pattern & Vulnerability & \textbf{exploits} \\ \hline
Course-Of-Action & Attack-Pattern & \multirow{2}{*}{\textbf{mitigates}} \\ \cline{1-2}
Course-Of-Action & Vulnerability &  \\ \hline
\end{tabular}
\caption{Relationships matrix}
\label{tbl:relationships}
\end{table}

\textbf{Object Extension:} All used objects follow the STIX 2.1 Specification but some have been extended with additional attributes/properties to cover the needs of CESO, as shown in Table \ref{tbl:obj_ext}.

\begin{table*}[th]
\centering
\begin{tabular}{|l|l|l|p{2.5in}|}
\hline
\textbf{Object} & \textbf{Attribute added} & \textbf{Type} & \textbf{Description} \\ \hline
Course-of-action & Difficulty (optional) & Integer & An integer from 1 to 5 (1 being easy and five being hard) declaring how difficult a course of action is evaluated to be executed by the player to resolve an incident. \\ \hline
Grouping & Scenario (mandatory) & String & A description that provides more details and context about the exercise scenario. \\ \hline
Identity & Recipient\_Group (optional) & String & The name of the recipient group in which players are split to receive different injects. \\ \hline
\end{tabular}
\caption{Objects Extension Matrix}
\label{tbl:obj_ext}
\end{table*}

\section{Automated Generation of Cybersecurity Exercise Scenarios}
\begin{figure}[th]
    \centering
    \includegraphics[width=.8\linewidth]{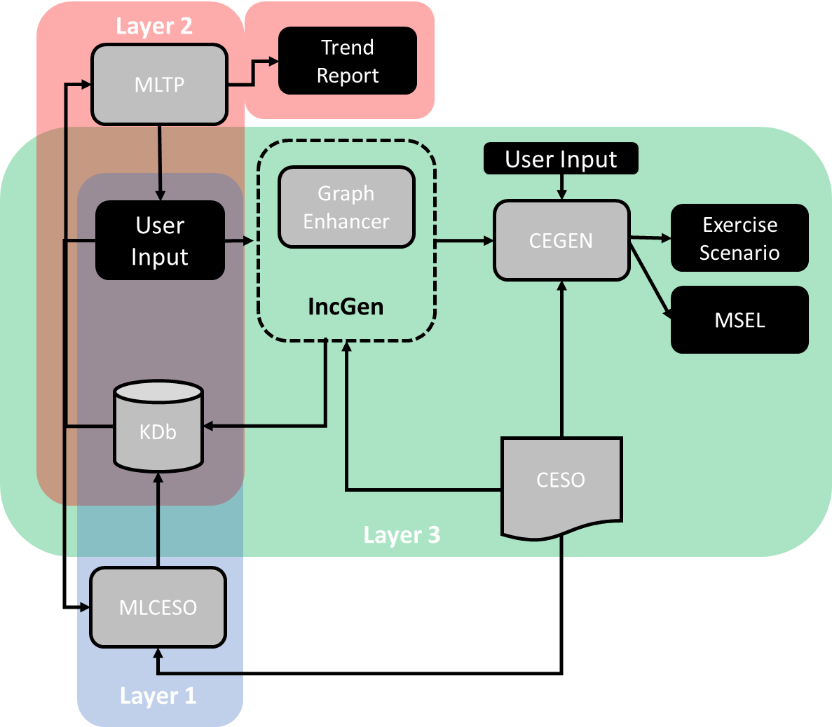}
    \caption{High-level overview of AiCEF.}
    \label{fig:aicef}
\end{figure}

To create the envisioned ML-powered Exercise Generation Framework, we opted to use Python and develop a set of tools that would perform a set of individual tasks, in the form of steps, which would help an EP, regardless of her experience, to create a timely and targeted Cybersecurity Exercise Scenario. The proof of concept framework we developed is AiCEF, and its general outline is illustrated in Figure \ref{fig:aicef}.

More concretely, to generate a concrete CSE scenario using AiCEF, the EP must perform the following steps. Initially, if the EP can generate a Trend Report on specific tags (e.g. Ransomware). In AiCEF, this is done through the MLTP module. Then, the EP provides a set of relevant articles or free text to be parsed and converted to Incident Breadcrumbs, as we call them in the implementation, using the MLCESO module. In essence, MLCESO takes as input a text and maps it to our CESO ontology. Therefore, this step extracts all the relevant information for a cybersecurity exercise from the text. The breadcrumbs are then stored in a knowledge database called KDb, with a name tag for quick retrieval. Using IncGen, the EP generates several Incidents based on the provided meta tags. The EP may enhance these Incidents using the Graph Enhancer module to simulate known APT activity and, this way, fill in the possible missing information. The Incidents come with a set of Injects that can be edited on the fly. All Incidents are named and stored locally for later use. Finally, using CEGen, the EP can create a CSE scenario by defining various attributes like CSE name, number of Events and Incidents. The various objects are then merged into a single STIX 2.1 graph, and the scenario is generated along with a State of the World storyline(SoW).

In the following paragraphs, we detail these steps and modules, providing some examples.

\subsection{Machine Learning to CESO (MLCESO)}
The most important step in our methodology is the creation of the ML pipeline that will parse free text and extract objects in CESO, as defined in the previous section. To do so, we need to train our ML following a well-structured methodology consisting of three phases: Corpus Building, Corpus Annotation, and Corpus Evaluation using NER.

\subsubsection{Corpus Building}
As shown in Table \ref{tbl:sources}, four Incident Sources have been identified as the initial input to our corpus. All these websites cover a wide variety of cyber security incidents in article format that date many years in the past. For simplicity, in this work, we decided to collect incidents from 2020-01 till 2022-03, which accounts for 2000 articles. All relevant articles were collected through automated web scraping.

\begin{table}[th]
\centering
\begin{tabular}{|l|l|}
\hline
\textbf{Webpage} & \textbf{Articles} \\ \hline
\texttt{bleepingcomputer.com} & 1000 \\ \hline
\texttt{securityaffairs.co} & 150 \\ \hline
\texttt{zdnet.com} & 350 \\ \hline
\texttt{databreaches.net} & 500 \\ \hline
Total & \textbf{2000} \\ \hline
\end{tabular}
\caption{Corpus Collection Count}
\label{tbl:sources}
\end{table}

\begin{figure*}[th]
    \centering
    \includegraphics[width=\linewidth]{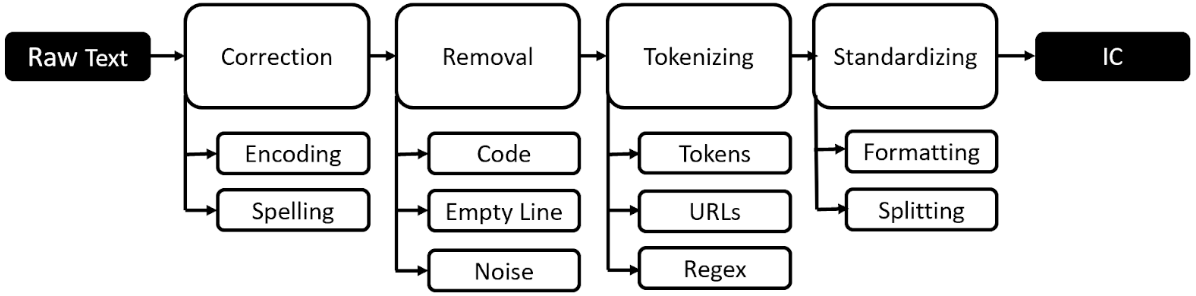}
    \caption{The Incident Corpus (IC) processing workflow.}
    \label{fig:incident_corpus}
\end{figure*}

Then, the raw text was processed using NLP techniques to form a reduced Incidents Corpus (IC). We used the text processing workflow illustrated in Figure \ref{fig:incident_corpus} to prepare the collected Incidents. Initially, all text was converted to the UTF-8 encoding scheme. Using dictionaries and the Textblob library\footnote{\url{https://github.com/sloria/TextBlob}}, we performed spelling corrections and removed special characters. Empty lines, specific stopwords and specific punctuation marks were removed using traditional NLP libraries like NLTK\footnote{\url{https://www.nltk.org/}} and spaCy \footnote{\url{https://spacy.io/}}. Moreover, all HTML or other programming codes, URLs, and paths were removed. Any illegal characters were also stripped, and all text was transformed to lowercase.

The standard Penn Treebank \cite{macintyrepenn} tokenisation rules were utilised for sentence tokenisation, and finally, standardisation processes were applied to tune the Incidents Text to facilitate annotation. At the end of this step, a corpus composed of Incidents was formed. As discussed, the corpus, from now on referred to as IC, contains 2000 cyber security articles. This accounts for 35.745 sentences containing 819.690 words leading to a vocabulary of 24594 terms. An example of a corpus line ready for annotation is the following:
\begin{lstlisting}[language=json,numbers=none]
{"text":"revil sodinokibi ransomware targets chinese users with dhl spam"}
\end{lstlisting}

\subsubsection{Corpus Annotation}
Following the CESO ontology, a simple model was developed comprising six steps to represent the annotation task. Entities and interconnections were formally described to align the efforts of converting words to tags in an Annotators Reference Document (ARD). This file, along with the corpus guidelines and CESO ontology, was given to the annotators to perform the annotation task using Prodigy\footnote{\url{https://prodi.gy}}. After completing the annotation, an inter-annotator agreement assessment took place using Cohen’s Kappa metric, and the gold standard version of the IC was finally produced.

Our annotation methodology consists of the following steps.

\noindent\textbf{Step 1 - Setting the Annotation Objectives:}
The main annotation objective was to create the appropriate semantic target to facilitate IC recognition by assigning the correct tag to in-context words in a sentence. Labelling all related words or sequences of words or text spans in the Cyber Incident context was crucial to perform efficient NER or text classification later. Each word or text span was labelled with a type identifier (tag) drawn from a vocabulary created based on the CESO ontology. It indicated what various terms denote in the context of a Cyber Incident and how they interconnect between them.

The following specific objective was set: Identify keywords, syntax, and semantic characteristics to detect i) Threat Actor ii) Cyber Security Incident or iii) Victim characteristics. If any of them is found, label them with the corresponding tag.

\noindent\textbf{Step 2 - Specifications Definition:} A concrete representation of the Annotation model to be used is created based on CESO.

An abstract model that practically represented the annotation objectives was defined. A three-category classification (Attacker, Attack, Victim) was introduced as the basis of this abstract model for identifying cyber-incident related terms in the text analysed. The category \textit{other} represents all remaining words out of context.

Our model M consists of a vocabulary of terms $T$, the relations between these terms $R$, and their interpretation $I$. Thus, our model can be represented as $M = < T, R, I >$ where:
\begin{itemize}
    \item T=\{CESO, Attacker, Attack, Victim, Other\}
    \item R=\{CESO:: = Attacker|Attack|Victim|Other\}
    \item I=\{Attacker= "list of attacker related terms in vocabulary",
    Attack =\textrm{"list of Cyber Security Incident or Attack terms in vocabulary"}, Victim = \{"list of victim-related terms in vocabulary"\}\\
    Other = \{"Other terms not related to the attacks"\}\}
\end{itemize}
\noindent\textbf{Step 3 - Annotator Reference Doc}: ARD is produced to help annotators in element identification and element association with the appropriate tags. The tags in Table \ref{tbl:Annotation_Tags} have been identified and mapped accordingly.

\begin{table}[th]
\centering
\begin{tabular}{|l|l|l|}
\hline
\textbf{Category} & \textbf{Tag} & \textbf{Link to CESO \& STIX 2.1} \\ \hline
\multirow{3}{*}{Attacker} & ATTACKER\_TYPE & Threat Actor Attribute \\ \cline{2-3}
 & ATTACKER\_NAME & Threat Actor Attribute, Identity \\ \cline{2-3}
 & ATTACKER\_ORIGIN & Location \\ \hline
\multirow{4}{*}{Attack} & MALWARE\_TYPE & Malware Attribute \\ \cline{2-3}
 & MALWARE\_NAME & Malware Attribute \\ \cline{2-3}
 & ATTACK\_TYPE (TECHNIQUE) & Attack Pattern \\ \cline{2-3}
 & VULNERABILITY & Vulnerability \\ \hline
\multirow{3}{*}{Victim} & SECTOR & Identity Attribute, Scenario \\ \cline{2-3}
 & ASSETS & Threat Actor Attribute \\ \cline{2-3}
 & TECHNOLOGY & Tool \\ \hline
\end{tabular}
\caption{Annotation Tags per Category}
\label{tbl:Annotation_Tags}
\end{table}

\noindent\textbf{Step 4 - Annotation Task: the annotation process is performed}

The annotation task aimed to label the words of IC corpus based on their semantic and syntactic characteristics. Two cybersecurity experts were assigned to label the words based on their semantic characteristics. By annotating the semantic characteristics of the words, the background information in each sentence was linked with CESO. The syntactic characteristics of the words were labelled using Prodigy. Table \ref{tbl:Annotation_Tags_EX} presents the annotation in action through some examples.

\begin{table*}[th]
\centering
\begin{tabular}{|l|lll|}
\hline
\textbf{Category} & \textbf{Tag} & \multicolumn{1}{l|}{\textbf{Annotator A-Tags}} & \textbf{Annotator B - Tags} \\ \hline
TEXT & \multicolumn{3}{l|}{qbot malware dropped via context aware phishing campaign infects the energy sector.}\\
 & \multicolumn{3}{l|}{russian hacking group claims 1000 windows machines compromised.} \\ \hline
\multirow{3}{*}{Attacker} & \multicolumn{1}{l|}{ATTACKER\_TYPE} & \multicolumn{1}{l|}{Hacking group} &  \\ \cline{2-4}
 & \multicolumn{1}{l|}{ATTACKER\_NAME} & \multicolumn{1}{l|}{} &  \\ \cline{2-4}
 & \multicolumn{1}{l|}{ATTACKER\_ORIGIN} & \multicolumn{1}{l|}{Russian} & Russian \\ \hline
\multirow{4}{*}{Attack} & \multicolumn{1}{l|}{MALWARE\_TYPE} & \multicolumn{1}{l|}{malware} & malware \\ \cline{2-4}
 & \multicolumn{1}{l|}{MALWARE\_NAME} & \multicolumn{1}{l|}{qbot} & qbot \\ \cline{2-4}
 & \multicolumn{1}{l|}{ATTACK\_TYPE (TECHNIQUE)} & \multicolumn{1}{l|}{phishing campaign} & phishing \\ \cline{2-4}
 & \multicolumn{1}{l|}{VULNERABILITY} & \multicolumn{1}{l|}{} &  \\ \hline
\multirow{3}{*}{Victim} & \multicolumn{1}{l|}{SECTOR} & \multicolumn{1}{l|}{energy} & energy \\ \cline{2-4}
 & \multicolumn{1}{l|}{ASSETS} & \multicolumn{1}{l|}{windows machines} & windows machines \\ \cline{2-4}
 & \multicolumn{1}{l|}{TECHNOLOGY} & \multicolumn{1}{l|}{windows} & windows \\ \hline
\end{tabular}
\caption{Annotation Tags per Category Example}
\label{tbl:Annotation_Tags_EX}
\end{table*}

\noindent\textbf{Step 5-Golden Standard Creation: the final version of the annotated Incident corpus is generated.}

The inter-annotator agreement (IAA) was validated using Cohen’s Kappa [40]\cite{10.1086/268657}. The formula used is defined as follows:
\begin{equation}
    \label{eq:kappa}
    k=\frac{p_0-p_e}{1-p_e}
\end{equation}
where $p_0$ expresses the relative observed agreement and $p_e$ the hypothetical probability of chance agreement.

The produced IC corpus has $N = 24594$ terms and $m = 4$ categories, and both annotators (A and B) agreed for the Attacker category 397 times, for the Attack category 1722 times, for the Victim 932 times and for the Irrelevant 21416.

Table \ref{tbl:Consistency_Matrix} shows the contingency matrix where each $x_{ij}$ represents the multitude of terms that annotator A classified in category i, but Annotator B is classified in category j, with i,j = 1,2,3,4. The proportions on the diagonal ($x_{ii}$) represent the proportion of terms in each category for which the two annotators agreed on the assignment.

\begin{table*}[th]
\centering
\begin{tabular}{|l|l|llll|l|}
\hline
\textbf{Annotator} &  & \multicolumn{4}{l|}{B} & \textbf{Total} \\ \hline
 & Category & \multicolumn{1}{l|}{Attacker} & \multicolumn{1}{l|}{Attack} & \multicolumn{1}{l|}{Victim} & Other & \textbf{} \\ \hline
\multirow{4}{*}{A} & Attacker & \multicolumn{1}{l|}{397} & \multicolumn{1}{l|}{10} & \multicolumn{1}{l|}{4} & 24 & \textbf{435} \\ \cline{2-7}
 & Attack & \multicolumn{1}{l|}{13} & \multicolumn{1}{l|}{1722} & \multicolumn{1}{l|}{8} & 9 & \textbf{1752} \\ \cline{2-7}
 & Victim & \multicolumn{1}{l|}{10} & \multicolumn{1}{l|}{2} & \multicolumn{1}{l|}{926} & 15 & \textbf{953} \\ \cline{2-7}
 & Other & \multicolumn{1}{l|}{16} & \multicolumn{1}{l|}{10} & \multicolumn{1}{l|}{12} & 21416 & \textbf{21454} \\ \hline
\textbf{Total} & \textbf{} & \multicolumn{1}{l|}{\textbf{436}} & \multicolumn{1}{c|}{\textbf{1744}} & \multicolumn{1}{l|}{\textbf{950}} & \textbf{21464} & \textbf{24594} \\ \hline
\end{tabular}
\caption{Consistency Matrix}
\label{tbl:Consistency_Matrix}
\end{table*}

The observed agreement $p_o$ is:
\[
p_o=\frac{397+1722+926+21416}{24594}=0,996
\]
and the expected change agreement, thus the proportion of terms which would be expected to agree by chance is:
\[p_e=\frac{\frac{436\times 435}{24594}+\frac{1744\times 1752}{24594}+\frac{950\times 953}{24594}+\frac{21464\times 21454}{24594}}{24594}\\=0,768 (76,8\%)
\]
so, according to Equation \ref{eq:kappa} the Cohen's Kappa is $k=\frac{p_0-p_e}{1-p_e}=\frac{0,228}{0,232}= 0,98$. Thus, based the Cohen’s kappa value of 0.98, we can safely conclude \cite{wilhelmson2011handbook} that the level of agreement for the corpus annotation task was almost perfect.

\subsubsection{Training \& Evaluation Using NER}
The following methodology has been used to train and evaluate our Named Entity Recognition (NER) agent.
\begin{enumerate}
    \item \textbf{Preprocessing:} The corpus has already been annotated, with each line of the corpus stored as a list of token-tag pairs. Each token was represented by a word embedding using the pre-trained English language model of the spaCy NLP library.
    \item \textbf{Build a model using spaCy}
    \item \textbf{Training:} To train the models in spaCy, we specified a loss function for the model to measure the distance between prediction and truth, and a batch-wise gradient descent algorithm was specified for optimisation.

One NER model was trained per object as presented in Table \ref{tbl:AI_Models_Scores}.

During the process that can be summarised in the flow above, several decisions were made to further improve the accuracy by retraining the models.  Since the aim was to create NER models that reach an F1 Score of $\approx\%$80\%, we iteratively extended the annotation and trained the model to pass this threshold.  


\item \textbf{Evaluation:} The performance assessment of the model was conducted by applying the model to the preprocessed validation data.
\end{enumerate}
\begin{table*}[th]
\centering
\begin{tabular}{|l|l|lll|}
\hline
\textbf{Category} & \textbf{Tag} & \multicolumn{1}{l|}{\textbf{Presicion}} & \multicolumn{1}{l|}{\textbf{Recall}} & \textbf{F1} \\ \hline
\multirow{3}{*}{Attacker} & ATTACKER\_TYPE & \multicolumn{1}{l|}{100.00} & \multicolumn{1}{l|}{83.33} & 90.11 \\ \cline{2-5}
 & ATTACKER\_NAME & \multicolumn{1}{l|}{95.29} & \multicolumn{1}{l|}{87.10} & 91.01 \\ \cline{2-5}
 & ATTACKER\_ORIGIN & \multicolumn{3}{l|}{Used Native Spacy LOC tag (no training)} \\ \hline
\multirow{4}{*}{Attacker} & MALWARE\_TYPE & \multicolumn{1}{l|}{80.56} & \multicolumn{1}{l|}{76.32} & 78.38 \\ \cline{2-5}
 & MALWARE\_NAME & \multicolumn{1}{l|}{95.29} & \multicolumn{1}{l|}{87.10} & 91.01 \\ \cline{2-5}
 & ATTACK\_TYPE (TECHNIQUE) & \multicolumn{1}{l|}{88.60} & \multicolumn{1}{l|}{87.07} & 87.83 \\ \cline{2-5}
 & VULNERABILITY & \multicolumn{1}{l|}{87.50} & \multicolumn{1}{l|}{84.00} & 85.71 \\ \hline
\multirow{3}{*}{Victim} & SECTOR & \multicolumn{1}{l|}{85.84} & \multicolumn{1}{l|}{84.07} & 84.95 \\ \cline{2-5}
 & ASSETS & \multicolumn{1}{l|}{87.02} & \multicolumn{1}{l|}{89.06} & 88.03 \\ \cline{2-5}
 & TECHNOLOGY & \multicolumn{1}{l|}{87.60} & \multicolumn{1}{l|}{89.93} & 88.70 \\ \hline
\end{tabular}
\caption{AI models' Scores.}
\label{tbl:AI_Models_Scores}
\end{table*}

While the results seem satisfactory, one can achieve further performance improvements in some tags.

We made an extra evaluation step with two experts against a set of 100 articles not used before in the training or evaluation steps. The aim was to evaluate the models against the selected tags empirically. The two reviewers have scored the NER accuracy per tag as presented in Table \ref{tbl:models_vs_rev}:
\begin{itemize}
    \item \textbf{HIT:} The tag was correctly assigned or not.
    \item \textbf{PARTIAL:} The tag was correctly assigned or not, but not for all values
    \item \textbf{MISS:} The tag was either assigned wrongly or was not assigned at all when it should
\end{itemize}

\begin{table*}[th]
\centering
\begin{tabular}{|l|p{1.4in}|lll|lll|lll|l|}
\hline
Category & \textbf{TAG} & \multicolumn{3}{l|}{\textbf{Reviewer 1}} & \multicolumn{3}{l|}{\textbf{Reviewer 2}} & \multicolumn{3}{l|}{\textbf{Average}} & \textbf{F1} \\ \hline
 & SCORE TYPE & \multicolumn{1}{l|}{H} & \multicolumn{1}{l|}{P} & M & \multicolumn{1}{l|}{H} & \multicolumn{1}{l|}{P} & M & \multicolumn{1}{l|}{H} & \multicolumn{1}{l|}{P} & M &  \\ \hline
\multirow{4}{*}{Attacker} & ATTACKER\_TYPE & \multicolumn{1}{l|}{90} & \multicolumn{1}{l|}{5} & 5 & \multicolumn{1}{l|}{90} & \multicolumn{1}{l|}{5} & 5 & \multicolumn{1}{l|}{90} & \multicolumn{1}{l|}{5} & 5 & 90.11 \\ \cline{2-12}
 & ATTACKER\_NAME & \multicolumn{1}{l|}{63} & \multicolumn{1}{l|}{20} & 17 & \multicolumn{1}{l|}{59} & \multicolumn{1}{l|}{24} & 17 & \multicolumn{1}{l|}{61} & \multicolumn{1}{l|}{22} & 17 & 91.01 \\ \cline{2-12}
 & ATTACKER\_ORIGIN & \multicolumn{1}{l|}{70} & \multicolumn{1}{l|}{25} & 15 & \multicolumn{1}{l|}{65} & \multicolumn{1}{l|}{30} & 15 & \multicolumn{1}{l|}{67.5} & \multicolumn{1}{l|}{17.5} & 15 & 98 \\ \cline{2-12}
 & MALWARE\_TYPE & \multicolumn{1}{l|}{82} & \multicolumn{1}{l|}{11} & 7 & \multicolumn{1}{l|}{80} & \multicolumn{1}{l|}{12} & 8 & \multicolumn{1}{l|}{81} & \multicolumn{1}{l|}{11.5} & 7.5 & 78.38 \\ \hline
\multirow{3}{*}{Attack} & MALWARE\_NAME & \multicolumn{1}{l|}{72} & \multicolumn{1}{l|}{14} & 24 & \multicolumn{1}{l|}{71} & \multicolumn{1}{l|}{14} & 25 & \multicolumn{1}{l|}{71.5} & \multicolumn{1}{l|}{14} & 24.5 & 91.01 \\ \cline{2-12}
 & ATTACK\_TYPE (TECHNIQUE) & \multicolumn{1}{l|}{84} & \multicolumn{1}{l|}{11} & 5 & \multicolumn{1}{l|}{86} & \multicolumn{1}{l|}{9} & 5 & \multicolumn{1}{l|}{85} & \multicolumn{1}{l|}{10} & 5 & 87.83 \\ \cline{2-12}
 & VULNERABILITY & \multicolumn{1}{l|}{75} & \multicolumn{1}{l|}{10} & 15 & \multicolumn{1}{l|}{79} & \multicolumn{1}{l|}{10} & 15 & \multicolumn{1}{l|}{77} & \multicolumn{1}{l|}{10} & 15 & 85.71 \\ \hline
\multirow{3}{*}{Victim} & SECTOR & \multicolumn{1}{l|}{84} & \multicolumn{1}{l|}{16} & 0 & \multicolumn{1}{l|}{86} & \multicolumn{1}{l|}{13} & 1 & \multicolumn{1}{l|}{85} & \multicolumn{1}{l|}{14.5} & 0.5 & 84.95 \\ \cline{2-12}
 & ASSETS & \multicolumn{1}{l|}{90} & \multicolumn{1}{l|}{7} & 3 & \multicolumn{1}{l|}{90} & \multicolumn{1}{l|}{8} & 2 & \multicolumn{1}{l|}{90} & \multicolumn{1}{l|}{7.5} & 2.5 & 88.03 \\ \cline{2-12}
 & TECHNOLOGY & \multicolumn{1}{l|}{90} & \multicolumn{1}{l|}{8} & 2 & \multicolumn{1}{l|}{86} & \multicolumn{1}{l|}{12} & 2 & \multicolumn{1}{l|}{88} & \multicolumn{1}{l|}{10} & 2 & 88.70 \\ \hline
\end{tabular}
\caption{AI Models Scores vs Reviewers Evaluation. H: Hit, P: Partial, M: Miss}
\label{tbl:models_vs_rev}
\end{table*}

The following findings should be highlighted:
\begin{enumerate}
    \item The hit rate of four (4) NER models has been identified as very weak, with an abnormal difference from the F1 score identified in the previous step.
    \item Names of Attackers or Malware can be a very vague topic to tackle using NER.
    \item The Attacker's Origin cannot be properly identified with the use of the out-of-the-box SpaCy LOC NER model. Locations are identified but can be related to the victim or are irrelevant to the attacker’s origin.
    \item The vulnerability NER model misses the correct formatting of CVE. This issue can be solved using a regex that accurately detects CVE in the text in combination with the model generated.
\end{enumerate}

\subsection{Incident Generation and Enhancement (INCGEN)}
\begin{figure*}[th]
    \centering
    \includegraphics[width=\linewidth]{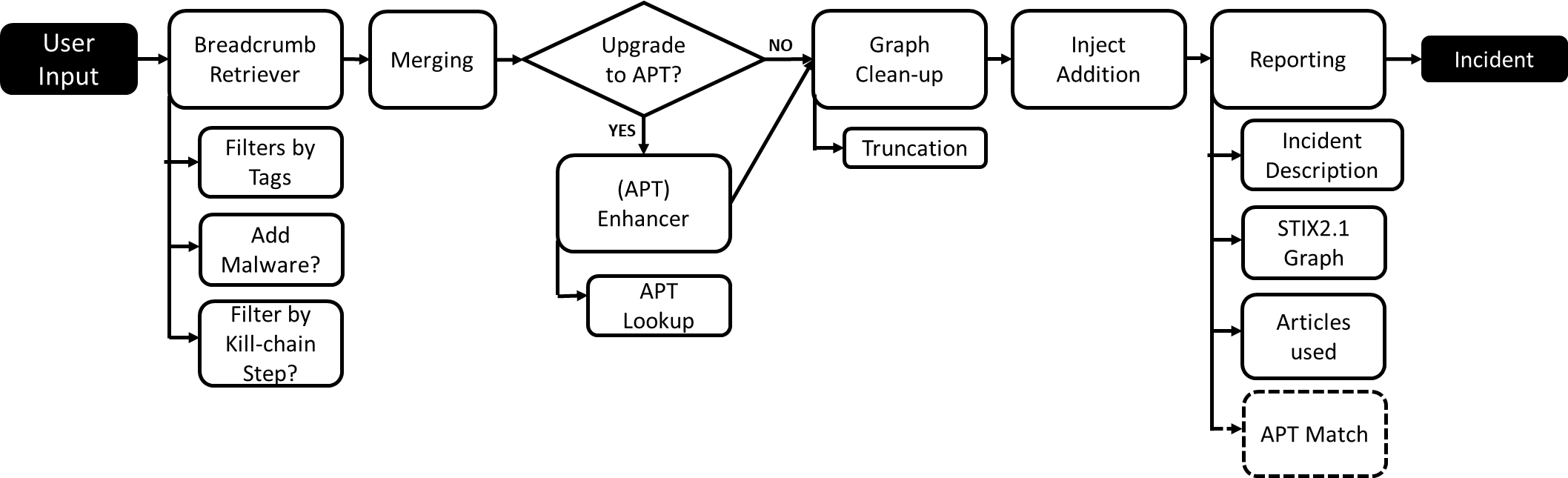}
    \caption{IncGen workflow}
    \label{fig:incgen_workflow}
\end{figure*}

Incident creation is the most important step of the scenario generation procedure and consists of several steps to achieve maximum customisation (Figure \ref{fig:incgen_workflow}). All of the steps can be automated, generating a variety of Incidents from which a Planner can choose to most fit.

The EP can choose to provide specific text or articles for conversion to Incidents or rely on a dynamic generation based on filtering parameters and a search of the existing database. Incidents can be enhanced with activity simulating TTPs of known APT actors.

\begin{table}[th]
    \centering
    \begin{tabular}{|l|r|}
    \hline
         \textbf{Source} & \textbf{Count}  \\
         \hline
         \texttt{bleepingcomputer.com}&1368 \\
         \hline
         \texttt{securityaffairs.co}& 169\\
         \hline
         \texttt{zdnet.com}& 495\\
         \hline
         \texttt{databreaches.net}& 938 \\
         \hline
         \textbf{Total} & \textbf{2970}\\
         \hline
    \end{tabular}
    \caption{Knowledge DB content per Source}
    \label{tbl:article_sources}
\end{table}

First, we need a set of texts to use as a baseline to generate our scenarios. Obviously, these articles would be parsed to be mapped with CESO so that our modules can process them. To this end, we used the sources of Table \ref{tbl:article_sources} to generate our Knowledge database (KDb). Evidently, not all articles that one may send for parsing may contain enough or even relevant information to generate a CSE scenario. They may be very generic or not relevant at all to cybersecurity. Therefore, we have introduced a threshold system to calculate the \textit{maturity} of an article that is parsed and the NER extracted tags. Our scoring system, which goes from 0 to 185, is illustrated in Algorithm \ref{alg:maturity}. In our implementation, we have set a threshold of 50 to consider a text relevant for representing a standalone incident in AiCEF.

\begin{algorithm}[th]
\caption{Compute the maturity of a parsed text.}\label{alg:maturity}
\begin{algorithmic}
\Require Set of Tags $T$
\State $maturity \gets 0$
\If{$Attacker\_Type\in T$ OR $Attack\_Type\in T$}
\State $maturity \gets 50$
    \If{$Vulnerability\in T$}
         $maturity \gets maturity+15$
    \Else~  $maturity \gets maturity-10$
    \EndIf
    \If{$Malware\in T$} $maturity \gets maturity+15$
    \Else~ $maturity \gets maturity-10$
    \EndIf
    \If{$Attack\_Type\in T$}
        \State $maturity \gets maturity+15$
        \If{$Attacker\_Type\in T$}
            \State $maturity \gets maturity+50$
            \If{$Technology\in T$}
                 $maturity \gets maturity+10$
            \EndIf
            \If{$Sector\in T$}
                 $maturity \gets maturity+10$
            \EndIf
            \If{$Assets\in T$}
                 $maturity \gets maturity+10$
            \EndIf
            \If{$Attackers\_Origin\in T$}
                 $maturity \gets maturity+10$
            \EndIf
        \EndIf
    \EndIf
\EndIf
\State \Return $maturity$
\end{algorithmic}
\end{algorithm}

\begin{figure*}[th]
    \centering
    \includegraphics[width=\linewidth]{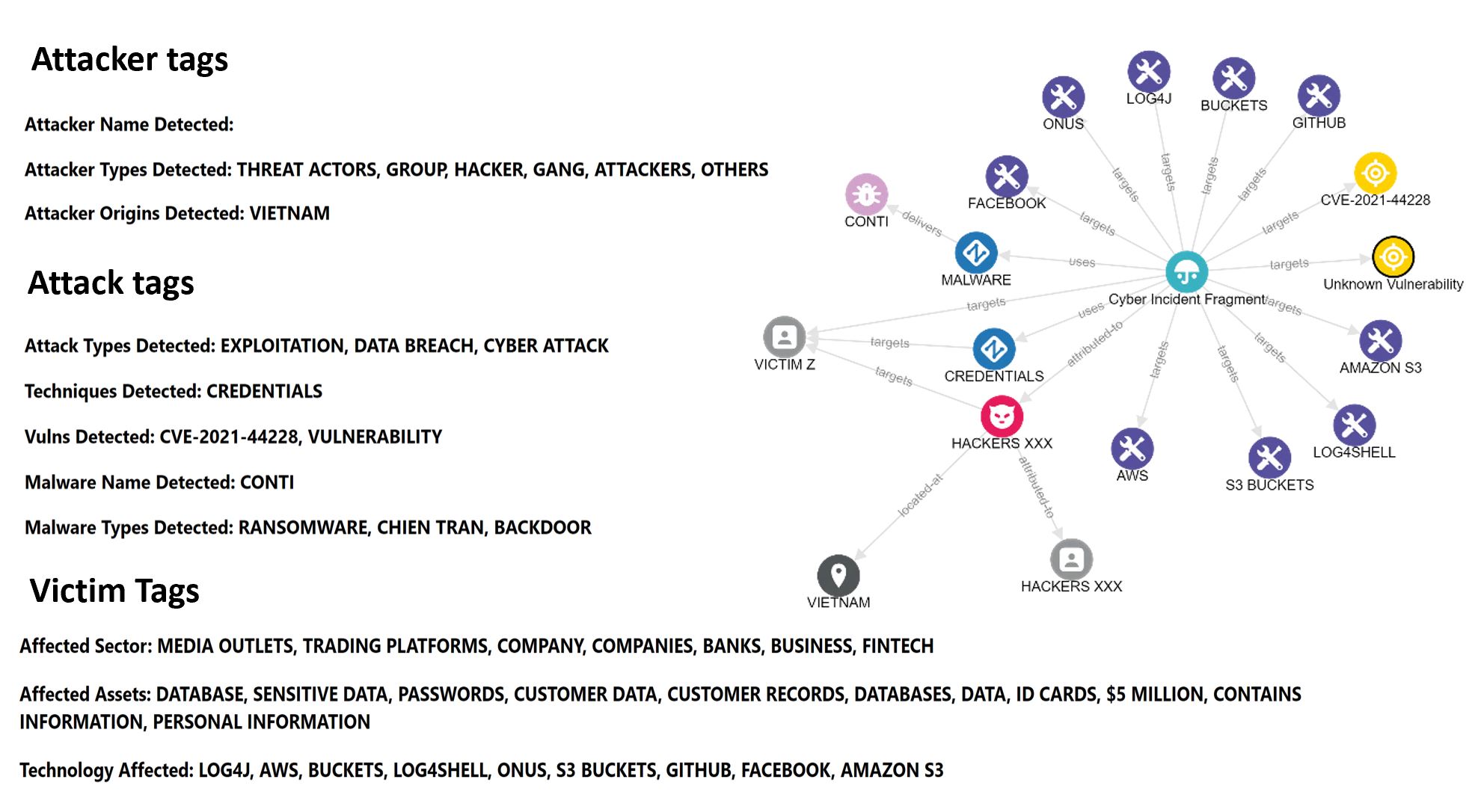}
    \caption{IncGen output report and visualisation}
    \label{fig:incgen_output}
\end{figure*}

Two types of enhancements were applied to improve the automatically NER exported tags, namely REGEX and Hard-coded groups of Strings. Thus, the following tags have been further enhanced:

\begin{itemize}
\item Attackers Name: NER + Hardcoded Groups of Strings from MITRE APT list \cite{MITRE}
\item Attackers Origin: No NER, Hardcoded Groups of Strings,
\item Malware Name: NER + Hardcoded Groups of Strings from MITRE APT list,
\item Technique: NER + Hardcoded Groups of Strings from MITRE APT list,
\item Vulnerability: NER + CVE REGEX.
\end{itemize}
The above enhancements greatly improved the tag detection rates, achieving almost 99\% in the Vulnerability tag. Moreover, based on the analysis of the most prominent extracted tags, the tag groups of Table \ref{tbl:topics} were assigned to the training topics meta tag to help categorise text for later use in an exercise scenario-building process.
An output report and visualisation (using stixview\footnote{https://github.com/traut/stixview} library) of IncGen utilising the improved MLCESO tag detection can be seen in Figure \ref{fig:incgen_output}.

\begin{table}[th]
    \centering
    \begin{tabular}{|l|p{3.25in}|}
    \hline
         \textbf{Training Topic}& \textbf{Tags} \\
         \hline
         INCIDENT HANDLING & MALWARE, RANSOMWARE, APT, CYBER, ATTACK, WEBSITE, HACKER, EXPLOIT, ZERO-DAY\\
         \hline GDPR & PRIVACY, DATA LEAKAGE, PERSONAL DATA, EXFILTRATION, CLOUD, SENSITIVE DATA, DATA, GOOGLE DRIVE, AWS, MEDICAL DATA, PASSPORT\\
            \hline CYBER HYGIENE &
            PASSWORD, ACCOUNT, USERNAME, LOGIN, ACCOUNTS, FILES, CREDENTIALS\\
            \hline PHISHING \& SOCIAL ENGINEERING&
            PHISHING, SCAM, FRAUD, VISHING, IMPERSONATION, BEC, EMAIL, GMAIL\\
            \hline SOCIAL MEDIA&
            FACEBOOK, TWITTER, LINKEDIN, META, INSTAGRAM\\
            \hline BYOD &
            MOBILE, ANDROID, IOS, LAPTOP, IOT, GOOGLE PLAY
            \\
            \hline

    \end{tabular}
    \caption{Training topics}
    \label{tbl:topics}
\end{table}

\subsection{APT Enhancer}

To simulate the activity of known APT groups basic STIX 2.1 structure was created per actor using the Groups from MITRE from which various attributes and TTPs were automatically extracted to populate our database. Thus, we generated a STIX 2.1 graph that can be used to compare and enhance other graphs. In this sense, during the enhancement process of an incident, the corresponding extracted graph is compared to all known APT actors and the most similar is proposed for enhancement. For each supported STIX 2.1 object type, the object similarity function (STIX 2.1 Python API) checks if the values for a specific set of properties match. Each matching property is separately weighted due to the fact that properties can have different levels of importance based on semantic similarity. The similarity score can range from 0 to 100, with 0 score representing no similarity between two objects compared.

In AiCEF, the EP may merge graphs completely or use only fragments. Thus, the EP may merge a draft incident graph with that of any known APT or proceed to merge with more than one APT actor's combined TTPs.
\subsection{Storyline Text Generation}

The Storyline Text Generator (STG) creates synthetic text based on predefined input. Using a Python text generator and Generative Pre-trained Transformer 2 (GPT-2)\footnote{\url{https://openai.com/blog/better-language-models/}}, AI large-scale unsupervised language model which can create coherent paragraphs of text from small pieces of text input.

\subsection{Trend Prediction Module (MLTP)}

The Trend Prediction Module provides valuable information to any EP by deep diving into the KDb and extracting interesting trends based on predetermined Training Objectives to compile a Trend Report.
To compile the report MLTP performs the following steps:
\begin{enumerate}
\item Step1. Receives Input like Filter Tags (Sector, Attack Type, Training Objective)
\item Step2. Extract incident statistics for the specified sector like Training Objective Breakdown, Top Attackers, Top Techniques, Top Malware used, Top Vulnerabilities used.
\item Step3. Perform Time Series Analysis on data for the Specific Attack Type and/or Training Objective, plotting and calculating Future Trends.
\end{enumerate}
In our implementation, we chose the SARIMA\footnote{SARIMA is Seasonal ARIMA, or simply put, ARIMA with a seasonal component. ARIMA is a statistical analysis model that uses time-series data to predict future trends} equation to represent the trends on the existing KDb of 2970 articles as represented in Table \ref{tbl:article_sources}.

\subsection{Cyber Exercise Generation (CEGEN)}
\begin{figure*}[th]
    \centering
    \includegraphics[width=\linewidth]{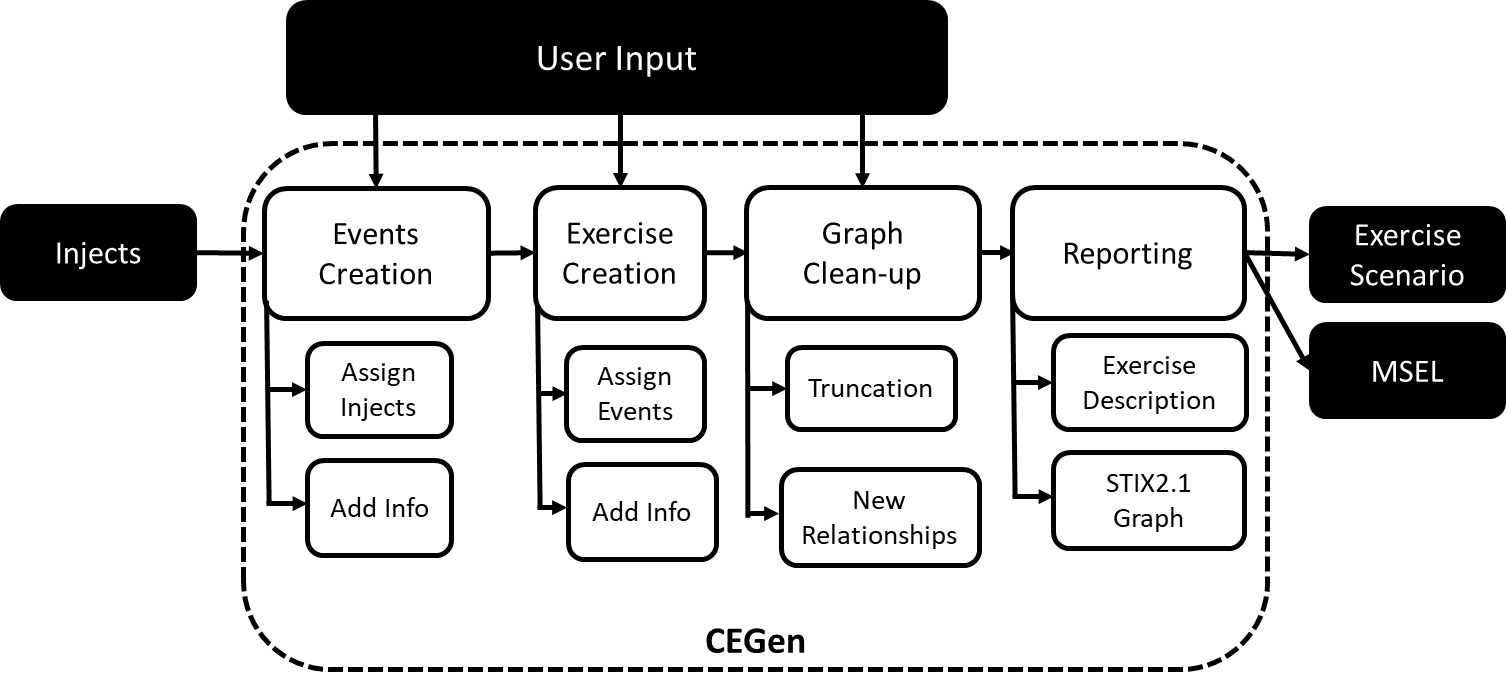}
    \caption{CEGen workflow}
    \label{fig:cegen_workflow}
\end{figure*}

The exercise generation flow can be broken down into several steps (Figure \ref{fig:cegen_workflow}).

Using the CEGEN module will lead to the generation of the following outputs:

\begin{itemize}
    \item  Exercise Scenario Summary along with a State of the World (SoW) skeleton
    \item  A MSEL tree skeleton (name and description)
\end{itemize}

\subsection{Putting everything together}
Now that we have described the main modules and their functionality, we may present AiCEF and present its outcomes. At some point, the EP will populate KDb with incidents which will be converted to graphs based on our ontology, CESO. Once the EP wishes to create a new scenario for a CSE, depending on the intended objectives, he/she would provide AiCEF with a set of keywords. To facilitate the planner work, AiCEF can initially generate a trend report that would allow the EP to identify trends relevant to the objectives at the time of the exercise execution. Based on the keywords, AiCEF will crawl its database for the most relevant articles and return a corresponding graph. From there, the EP can enhance the graph by merging it with that of known threat groups. The new graph can then be filtered according to the intended Cyber Kill Chain phases to be simulated.

\begin{figure*}[th]
    \centering
    \includegraphics[width=\textwidth]{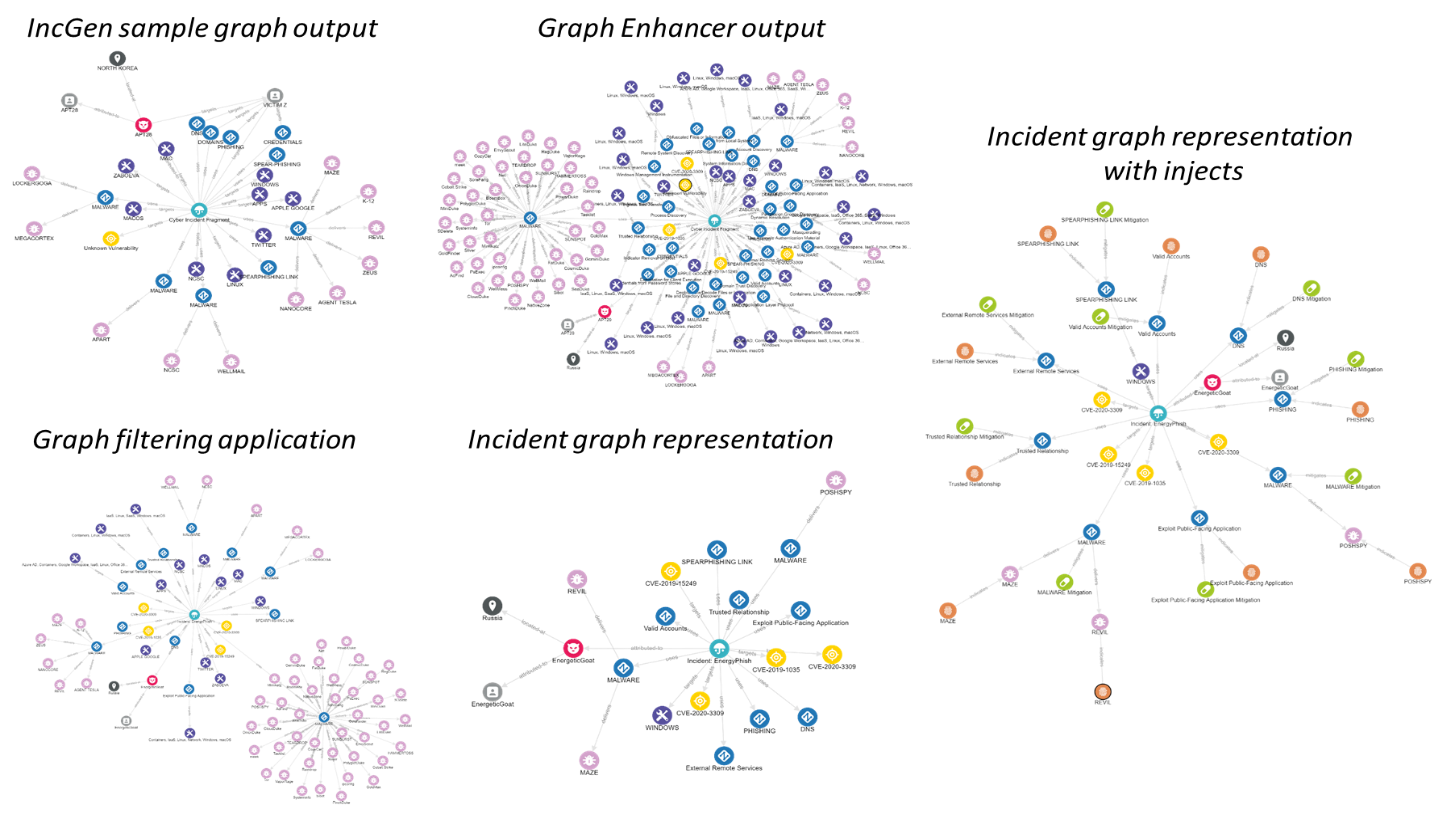}
    \caption{IncGen execution flow with intermediate representation steps}
    \label{fig:incgen_flow}
\end{figure*}

The above actions lead to a more specific incident graph representation that is ready to be populated with injects. A representation of the progress of an incident graph generation can be visualised in Figure \ref{fig:incgen_flow}.

\begin{figure*}[th]
    \centering
    \includegraphics[width=\textwidth]{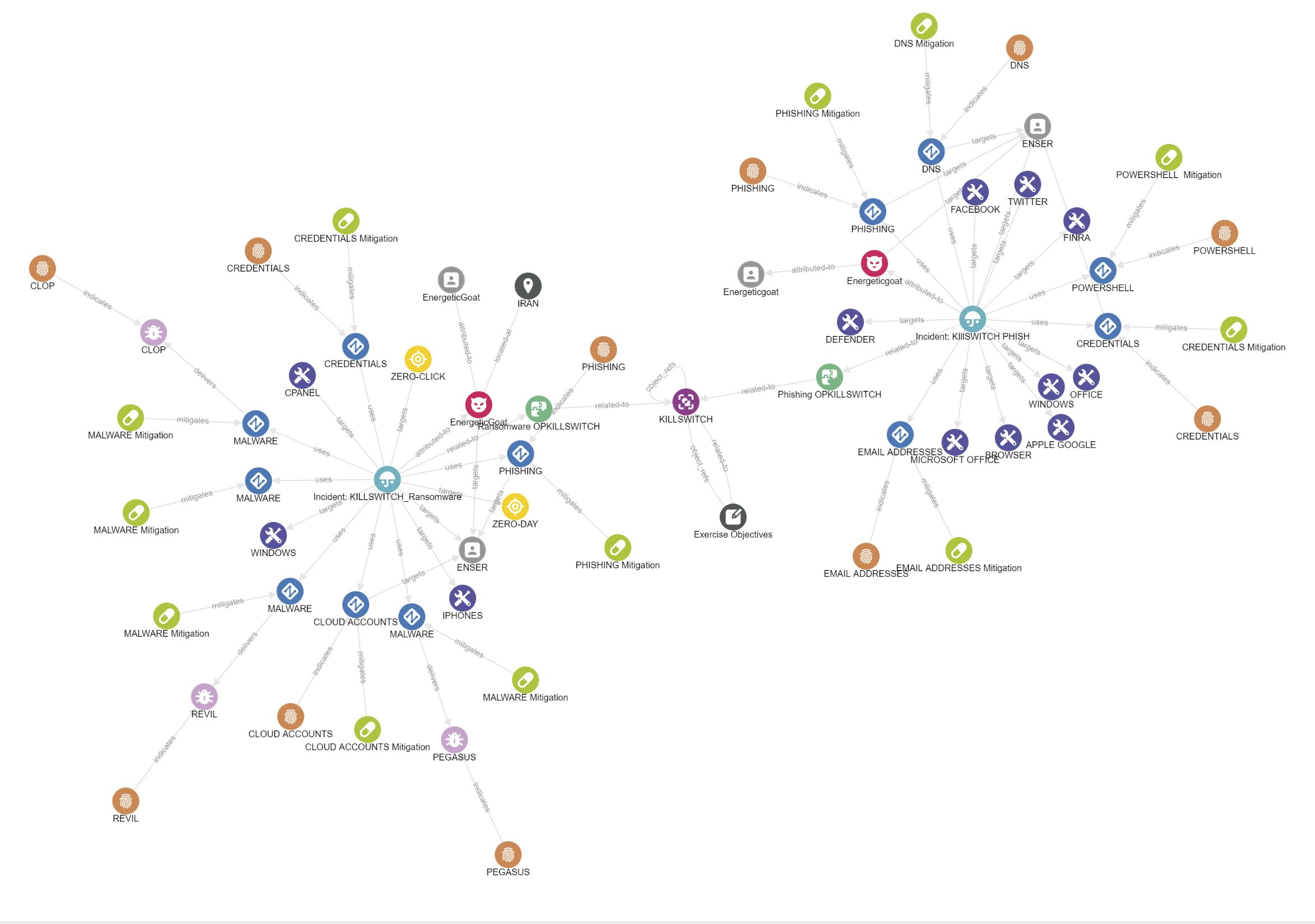}
    \caption{Sample exercise Graph visualisation}
    \label{fig:exgraph}
\end{figure*}

This process is repeated multiple times to generate the number of wanted incidents for a specific CSE. The EP follows the CEGen flow to compile a full exercise and generate a scenario (Figure \ref{fig:textgenerated}) and Exercise graph (Figure \ref{fig:exgraph}).
\begin{figure*}[th]

    \textbf{State of the World (SoW)}

    \fbox{\parbox{\textwidth}{
    Description: {\textbf{STG Input}:: ‘\textit{State sponsored Threat actor EnergeticGoat has historically targeted government organizations, non-government organizations (NGOs), think tanks, military, IT service providers, health technology and research, and telecommunications providers. With this latest attack, EnergeticGoat attempted to target the Energy Sector}’}

\textbf{STG Output:} “State sponsored Threat actor EnergeticGoat has historically targeted government organizations, non-government organizations (NGOs), think tanks, military, IT service providers, health technology and research, and telecommunications providers. With this latest attack, EnergeticGoat attempted to target the Energy Sector in South Europe. More specifically the following targets have been identified: 1) EU energy firms that have received investments from the Energysubsidiary, 2) SA government units and departments who receive funding from the Energysubsidiary, 3) the SA energy regulator, 4) the Energy Sector’s main lobbying and public relations organization, 5) the Energy Sector’s largest customer, the Energysubsidiary, and 6) the Energy Sector’s corporate regulator, the Energy Regulator of SA (ERA). Energy Sector entities and their customers and regulators are usually among the least likely organizations to be attacked by EnergeticGoat.

For example, the Energy Regulator of SA (ERA) is not a high-profile organization. It is a largely unregulated regulatory body that provides general oversight of the energy sector and serves as a key regulator and enforcement body of relevant energy laws in EU. Another example would be the Energy Corporation (Zesa). Zesa is a state-funded enterprise that provides essential infrastructure and related services to its citizens, including the electricity, gas, and broadband services that power large swaths in EU.

The attackers may have targeted other Energy Sector actors, such as Eskom or the SA Energy Institute (SAEI). However, it is thought that EnergeticGoat is specifically targeting Zesa and other organizations as a result of the threat actor’s perceived inability to target the SA Energy Sector as a whole. EnergeticGoat’s choice to target a South African energy firm with its latest attack is indicative of this perception.”
    }}

\textbf{Scenario Summary}

    \fbox{\parbox{\textwidth}{
    EXERCISE NAME: Energy Test

EXERCISE OBJECTIVES: Test detection of Attacks and Incident Handling Procedures

\#\#\# EVENT 1: EnergyWipeOut \#\#\#\
Description: {\textbf{STG Input}:: ‘\textit{State sponsored Threat Actor EnergeticGoat is believed to be behind cyber-attacks against Energy Corp. The attackers used PHISHING \& DATA DESTRUCTION techniques to}’}

\textbf{STG Output:} “State sponsored Threat Actor EnergeticGoat is believed to be behind cyber-attacks against Energy Corp. The attackers used PHISHING \& DATA DESTRUCTION techniques to infect the system with malware. In addition to the malware used, the attackers also used other tools to get around the security countermeasures of ENERGYCORP. It is understood that it will take at least a few days for the authorities to identify the attackers as they have only just informed the company about the attack. A spokesperson for the company said the company did not have any details as to the motive behind the attack as it was still investigating.

“We are working with law enforcement to bring this to the most serious level. The situation is under investigation and we cannot comment any further than that”

*** Incident 1.1: EnergyPhish ***
Description: {\textbf{STG Input}:: ‘\textit{EnergeticGoat deployed Phishing attacks against Energy Corp. Malicious word documents were sent}’}

\textbf{STG Output:}“EnergeticGoat deployed Phishing attacks against Energy Corp. Malicious word documents were sent to Energy Corp. employees in an attempt to obtain personal information. On October 26, 2022, a new employee of Energy Corp., received a suspicious email containing a malicious word document. The email identified as ‘malware’ was in fact a phishing attack with the subject ‘Vendors of Energy Corp. under Federal Investigation”}}
    \caption{Sample text of an AI-generated exercise.}
    \label{fig:textgenerated}
\end{figure*}

\section{Evaluation Methodology \& Results}
We developed a case study to help measuring the effectiveness of our proposed framework and underlying methodology. To this end, the steps below were followed.
\begin{enumerate}
    \item \textbf{Scenario Content Generation:} A group of exercise planners, of varying expertise, have been used to generate the same exercise scenario using traditional exercise means and the AiCEF methodology and tools while being monitored on timeliness, effectiveness, creativity and methodology used.
    \item \textbf{Content Evaluation:} The reports were then anonymised and given to a group of evaluators to grade the complexity, technical depth and richness of lessons learnt on the generated subset of exercise scenarios as per Objectives and KPIs set through a questionnaire.
    \item \textbf{Results collection and Analysis:} The results of this process were evaluated against the previously set KPIs to estimate:
    \begin{enumerate}
        \item Improved speed in Cyber Exercise Scenario generation (quantitative) using AiCEF.
        \item Improvement in quality in Cyber Exercise Scenario generation (qualitative) for inexperienced Planners using AiCEF.
        \item Improved relevance of proposed Cyber Exercise Scenarios to the current threat landscape (qualitative) using AiCEF.
    \end{enumerate}
\end{enumerate}

\subsection{Scenario Content Generation}
Four EPs were selected to individually generate a CSE scenario according to the provided high-level exercise requirements and specifications, see Figure \ref{fig:task_def}. The EPs were split into two groups based on their previous experience with the task. All EPs have deep knowledge of cyber security, and their skill sets resemble that of a CISO.
\begin{figure*}
    \fbox{\parbox{\textwidth}{
    Generate a CSE scenario for a cyber awareness exercise by filling in a provided Scenario Template. The CSE should include 2 events consisting of 1 incident each. All incidents should be accompanied by a short description of indicative injects to be sent to players. At least 3 inject descriptions per incident should be provided.

The company which will use the scenario is an \textbf{Energy Service Provider} and all its Employees can be potential Players.

 The exercise should last between 2-4 hours and can include technical artefacts for analysis.

The two main objectives of the exercise are:
\begin{enumerate}
    \item Provide awareness to employees regarding Phishing Attacks
    \item Provide awareness to employees regarding Ransomware Attacks
\end{enumerate}
These objectives can be updated and more can be added.
The Scenario Development task will be timed and should not last more than 4 hours
}}
    \caption{Task Definition}
    \label{fig:task_def}
\end{figure*}
Both groups consisted of one experienced and one inexperienced planner. The first group was briefly introduced to the basics of developing CSE scenarios, while the second one was provided a course on using AiCEF and the accompanying tools. Both groups were provided the same Scenario Template (ST) to fill in as an output of their task.

Then, we created a third group, consisting of Scripted Exercise Planner (SEP), using different parameters and flows of the AiCEF methodology and toolset.

The provided ST had the following generic structure:
\begin{itemize}
    \item Section 1: Storyline (SoW)
    \item Section 2: Scenario \& MSEL
    \item Section 3: Scenario Analysis
    \item Section 4: Resources Used
\end{itemize}
We provided detailed instructions of the expected content per paragraph to all involved planners to streamline the information of the generated reports and create homogeneous outputs to be evaluated in the later step.

As a result, five complete exercise scenarios were generated, as shown in Table \ref{tbl:scenarios}.

\begin{table*}[th]
\centering
\begin{tabular}{|l|l|p{0.75in}|l|l|l|}
\hline
\textbf{Eval\_Tag} & \textbf{Explanatory Name Tag} & \textbf{Exercise Expertise} & \textbf{AiCEF} & \textbf{Duration} & \textbf{Other Tools} \\ \hline
ExSc1 & Sc1:Exp(erienced)Hum(an) & YES & NO & 2h 00min & Google, MITRE \\ \hline
ExSc2 & Sc2:Nov(ice)Hum(an) & NO & NO & 2h 35min & Google \\ \hline
ExSc3 & Sc3:Exp(erienced)Hum(an)\&AI & YES & YES & 1h 20min & Online Resources \\ \hline
ExSc4 & Sc4:Nov(ice)Hum(an)\&AI & NO & YES & 2h 10 min & Google \\ \hline
ExSc5 & Sc5:AICEF & N/A & YES & 20min & - \\ \hline
\end{tabular}
\caption{Generated Scenarios details.}
\label{tbl:scenarios}
\end{table*}

\subsection{Scenario Content Evaluation}
To evaluate the scenarios above, we conducted an anonymous survey. To avoid bias, we invited a number of evaluators from different Cyber Awareness and Cyber Exercise Groups with varying expertise, ethnicity, and focus sectors to participate in the evaluation process. In total, 16 experts responded, whose demographic statistics are illustrated in Table \ref{tbl:demographics}.

\begin{table}[th]
\centering
\begin{subtable}[t]{0.45\linewidth}
\centering
\begin{tabular}{|l|r|}
\hline
\textbf{Countries} & \textbf{\#} \\ \hline
Greece & 2 \\ \hline
Austria & 1 \\ \hline
Italy & 1 \\ \hline
Belgium & 2 \\ \hline
Poland & 1 \\ \hline
Spain & 1 \\ \hline
Romania & 1 \\ \hline
Portugal & 1 \\ \hline
Czechia & 1 \\ \hline
Netherlands & 2 \\ \hline
France & 2 \\ \hline
Finland & 1 \\ \hline
\end{tabular}
\caption{Countries of origin of the experts}
\label{tbl:countries}
\end{subtable}
\begin{subtable}[t]{0.45\linewidth}
\centering
    \begin{tabular}{|l|r|}
    \hline
    \textbf{Sector} & \textbf{\#} \\ \hline
    Govermental & 6 \\ \hline
    Energy & 2 \\ \hline
    ICT & 4 \\ \hline
    Critical Infra. & 1 \\ \hline
    Law enforcement & 1 \\ \hline
    Education & 1 \\ \hline
    Other & 1 \\ \hline
    \end{tabular}
    \caption{Sector that experts are working in}
    \label{tbl:sector}
\end{subtable}
    \begin{subtable}[t]{\linewidth}
    \centering
    \begin{tabular}{|l|l|}
    \hline
    \textbf{Seniority} & \textbf{\#} \\ \hline
    Novice (small exercises) & 5 \\ \hline
    Medium (medium scale exercises for a few years) & 3 \\ \hline
    Expert (EU level, cross country exercises) & 4 \\ \hline
    Senior (large sized exercises) & 1 \\ \hline
    None & 3 \\ \hline
    \end{tabular}
    \caption{Seniority self-assessment.}
    \label{tbl:seniority}
    \end{subtable}
\caption{Demographics of the experts.}
\label{tbl:demographics}
\end{table}

The survey was in the form of an online questionnaire consisting of 11 questions. Eight questions were used to evaluate the generated Scenarios, two to be used as Turing Test to determine whether the AI used could be identified by humans and a set of complementary questions for demographic and future improvement purposes. All five scenarios were provided using only the ""Eval\_Tag" parameter for tracking purposes without providing additional information on the authors of the scenarios.

The eight scenario evaluation questions and their corresponding scores are the following:
\begin{enumerate}
    \item How do you evaluate the relevance of the State of the World text to the Objectives of the Exercise?
Score range: 0-4.
    \item How do you evaluate the relevance of the selected Events to the Objectives of the exercise?
Score range: 0-4.
    \item How do you evaluate the relevance of the selected Incidents to the Objectives of the exercise?
Score range: 0-4.
    \item How do you evaluate the Complexity of the Scenario?
Score range: 0-1.
    \item How do you evaluate the Technical Depth of the Scenario?
Score range: 0-2.
    \item How do you evaluate the Threat Actor’s description?
Score range: 1-3.
    \item How do you evaluate the used resources?
Score range: 0-2.
    \item Would you be willing to use this Scenario based on the task description?
Score range: 0-4.
\end{enumerate}
To evaluate the use of AI for exercise content generation, the following results were collected based on the questions below:
\begin{enumerate}
    \item How was the scenario generated?
    \item How skilled was the planner?
\end{enumerate}
Other questions revolved around the overall scenario development process:
\begin{enumerate}
    \item How much time did you invest in the Scenario Content Development?
    \item How do you define the scope/objectives of the exercise?
    \item How do you define the scenario content?
    \item What tools did you use to create the scenario or define the objectives if any?
\end{enumerate}
Finally, evaluators were asked to rank AI-powered tools as follows:
Rank the following AI-powered tools that could be created to support the design and implementation of future cyber exercises:
Automated extraction of Exercise Objects (Incidents, Injects) from unstructured information and DB storage
Lead generation for trend prediction of Training Topics
Automated Enrichment's of content to match realistic patterns and relationships of know Attackers
Automated Cyber Exercise Script/Scenario Generation

\subsection{Results Analysis}
The analysis of the input provided a good understanding of the strengths and potential areas for improvement of AiCEF. It also provided better insight into the exercise Scenario creation process, with good inputs for future improvement based on the experience of real EPs.
\begin{figure*}[th]
    \centering
    \includegraphics[width=.8\textwidth]{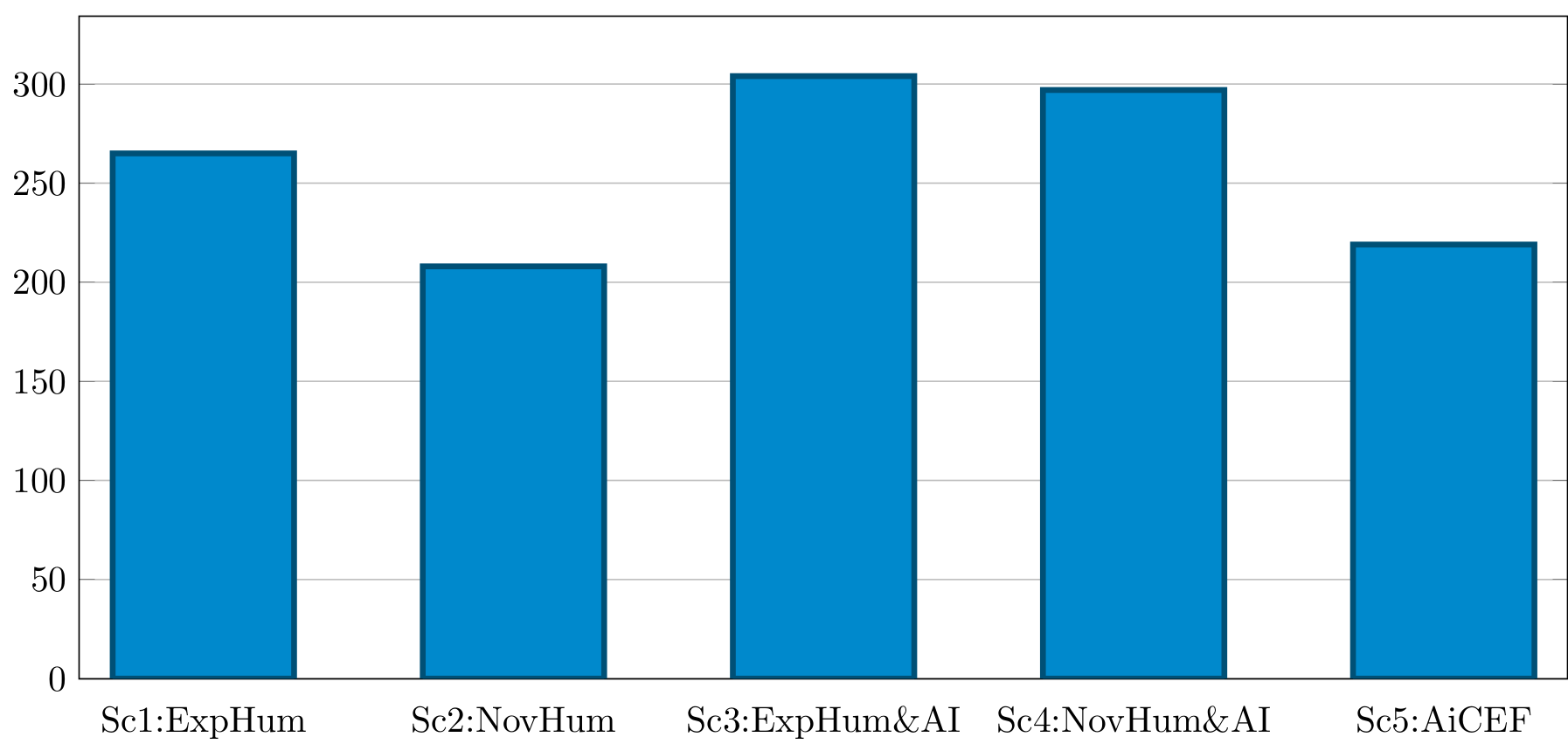}
    \caption{Overall performance of evaluated scenarios based on total score.}
    \label{fig:score}
\end{figure*}
Based on analysis of the provided input, we can safely conclude that both scenarios Sc3:ExpHum\&AI and Sc4:NovHum\&AI have scored higher than any other scenario with the help of AiCEF. Currently, the hybrid scenario generation approach of a human exercise planner using AiCEF outperforms a seasoned exercise planner, even when a planner is a novice. Furthermore, the Scripted Exercise Planner generated a relatively good Scenario (Sc5:AiCEF) that can be evaluated as equal, if not better, than that of a novice planner with strong Cyber Security background (SC2:NovHum).

\begin{figure*}[th]
    \centering
    \includegraphics[width=.8\textwidth]{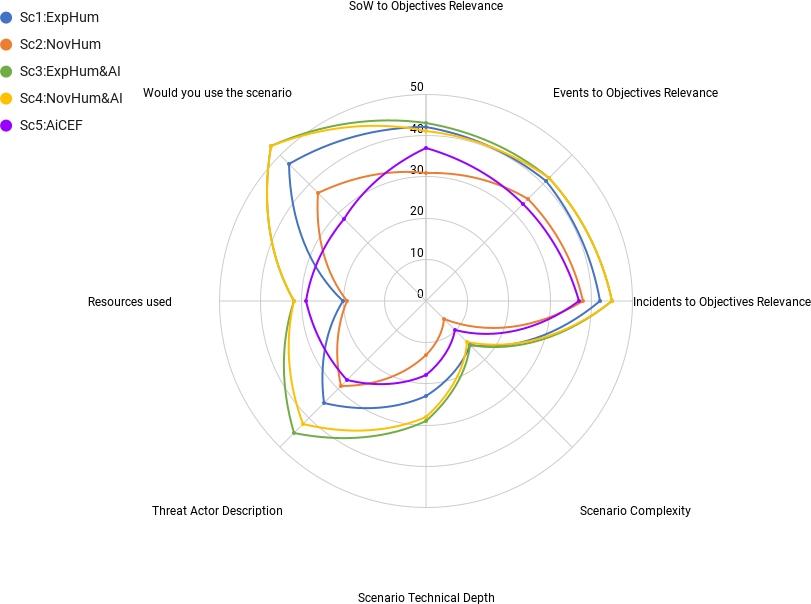}
    \caption{Scenario Evaluation Parameters}
    \label{fig:scenario_eval}
\end{figure*}

In what follows, we provide a breakdown of the parameters evaluated to highlight the strengths and weaknesses of using AiCEF based on the experts' input.

The use of AiCEF by a Scripted Exercise Planner performed well (top 3, outperforming humans) in \textit{Relevant Resources}, \textit{Events Relevance}, and \textit{Scenario Technical Depth}. On the other hand, AiCEF did not perform as well in the following aspects:
\textit{Threat Actor Description}, \textit{Scenario Complexity}, and \textit{Incidents to Objectives Relevance}. The above can be justified by the fact that the raw generated content can include conflicting information or content that might not match the high-level context requested. After human curation, the content can be easily improved to compete a seasoned exercise planner. In fact, AiCEF used by humans helped them excel in \textit{Scenario Creation}, dominating all categories versus their human counterparts. The human expert using AiCEF (Sc3:ExpHum\&AI) managed to create a better scenario 33,33\% faster than his expert peer using regular tools (Sc1:ExpHum).

\begin{figure*}[th]
    \centering
    \begin{subfigure}[t]{0.48\textwidth}
        \includegraphics[width=\linewidth]{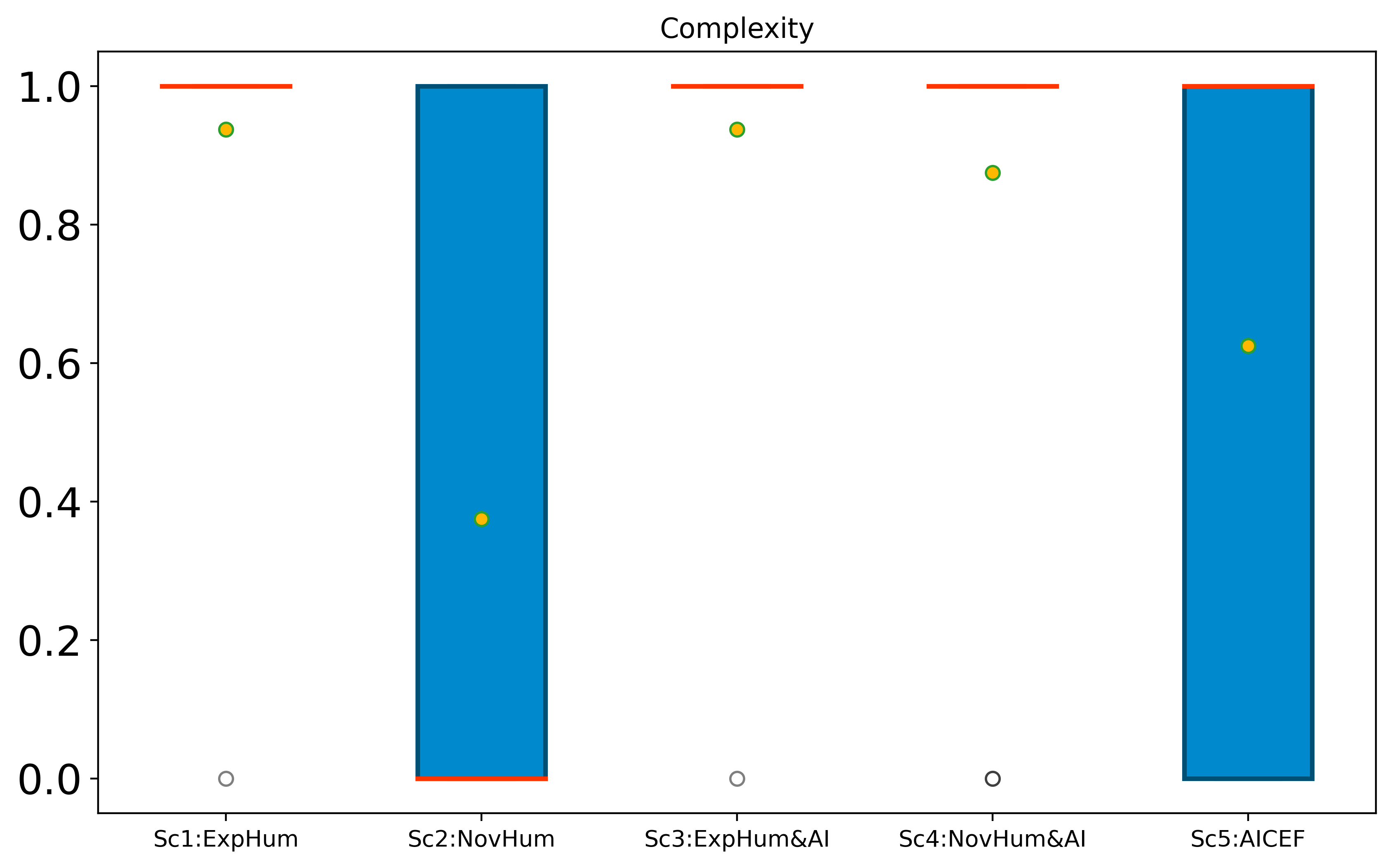}
        \caption{Complexity}
    \end{subfigure}
    \begin{subfigure}[t]{0.48\textwidth}
        \includegraphics[width=\linewidth]{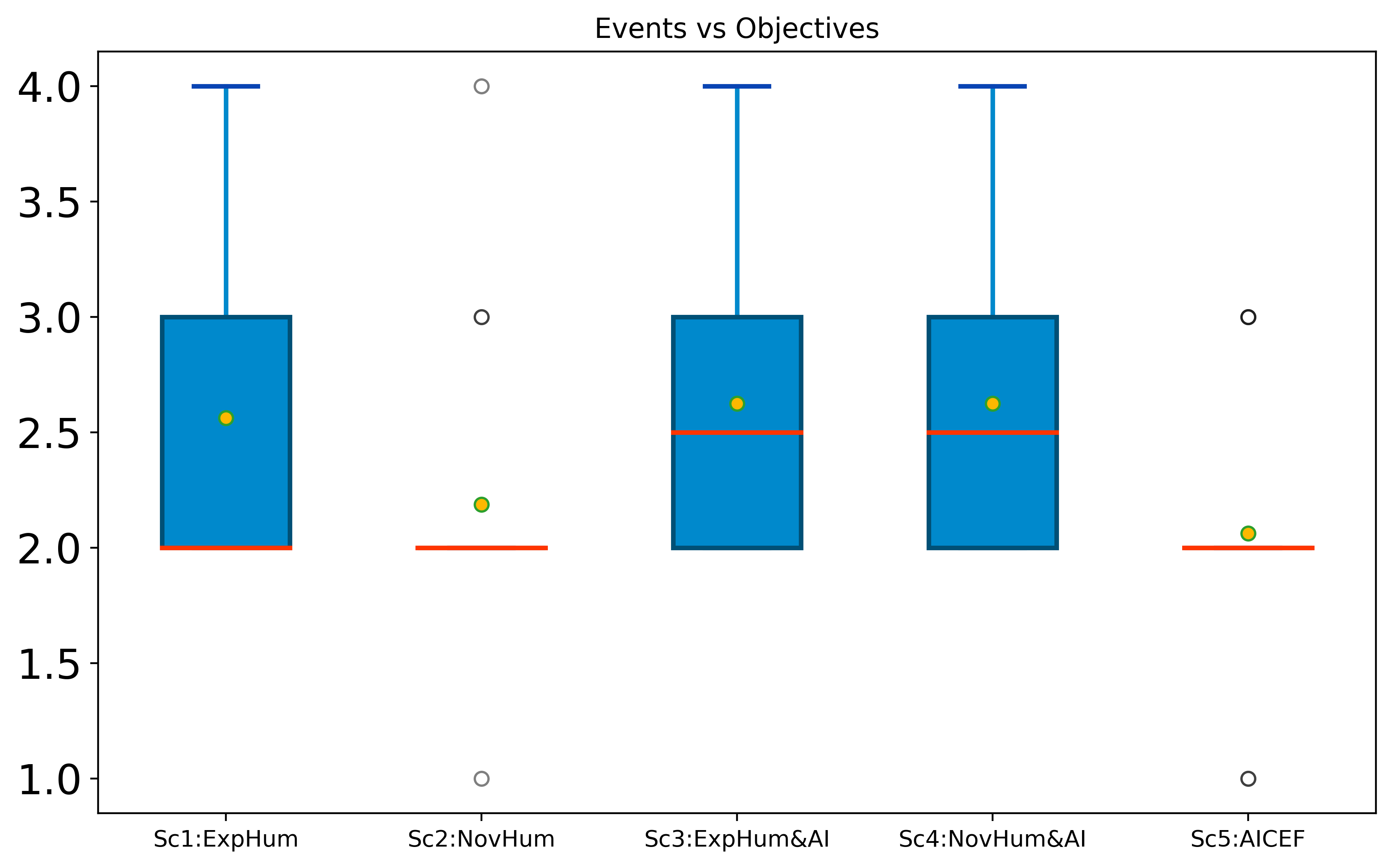}
        \caption{Events vs Objectives}
    \end{subfigure}

    \begin{subfigure}[t]{0.48\textwidth}
        \includegraphics[width=\linewidth]{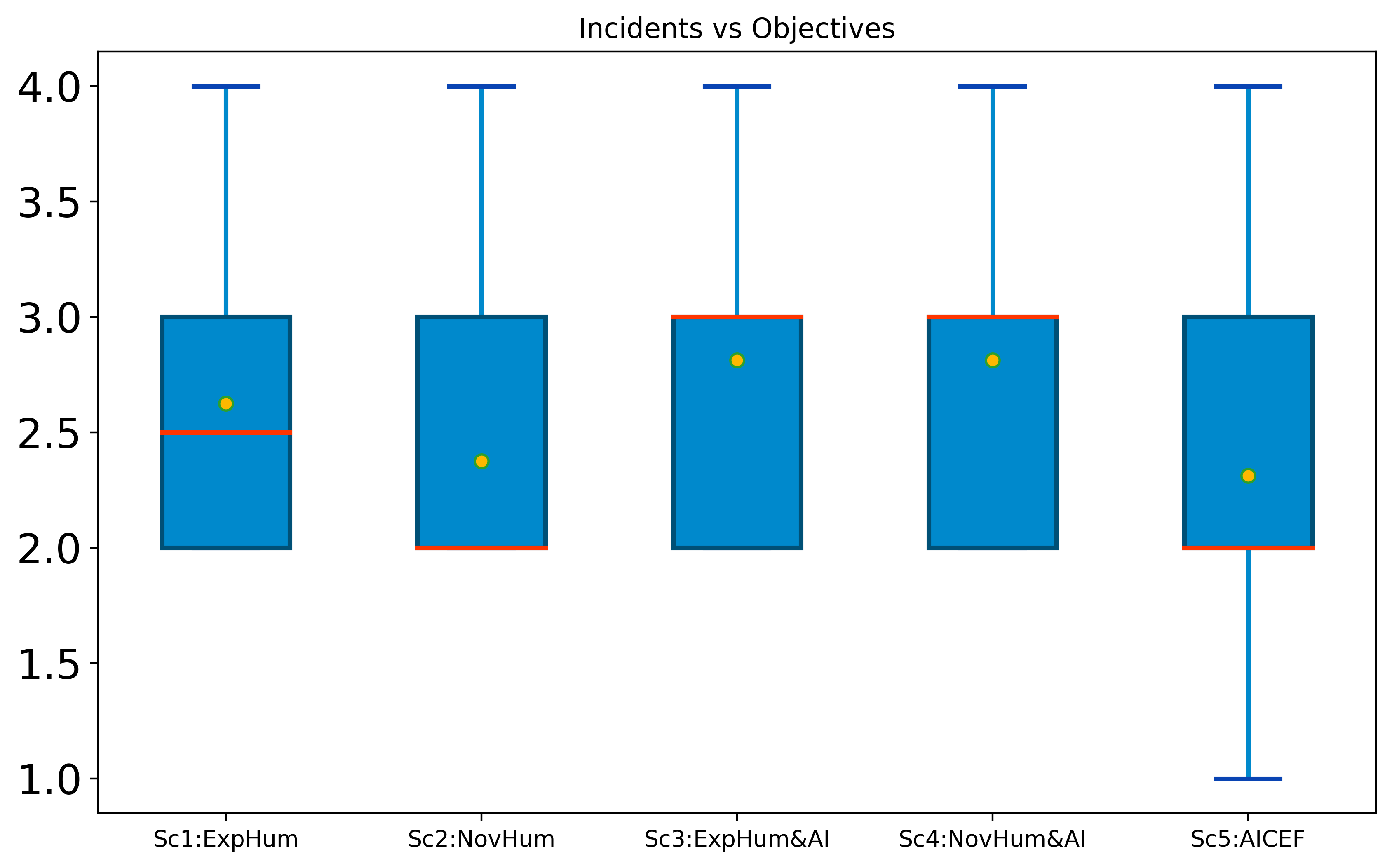}
        \caption{Incidents vs Objectives}
    \end{subfigure}
    \begin{subfigure}[t]{0.48\textwidth}
        \includegraphics[width=\linewidth]{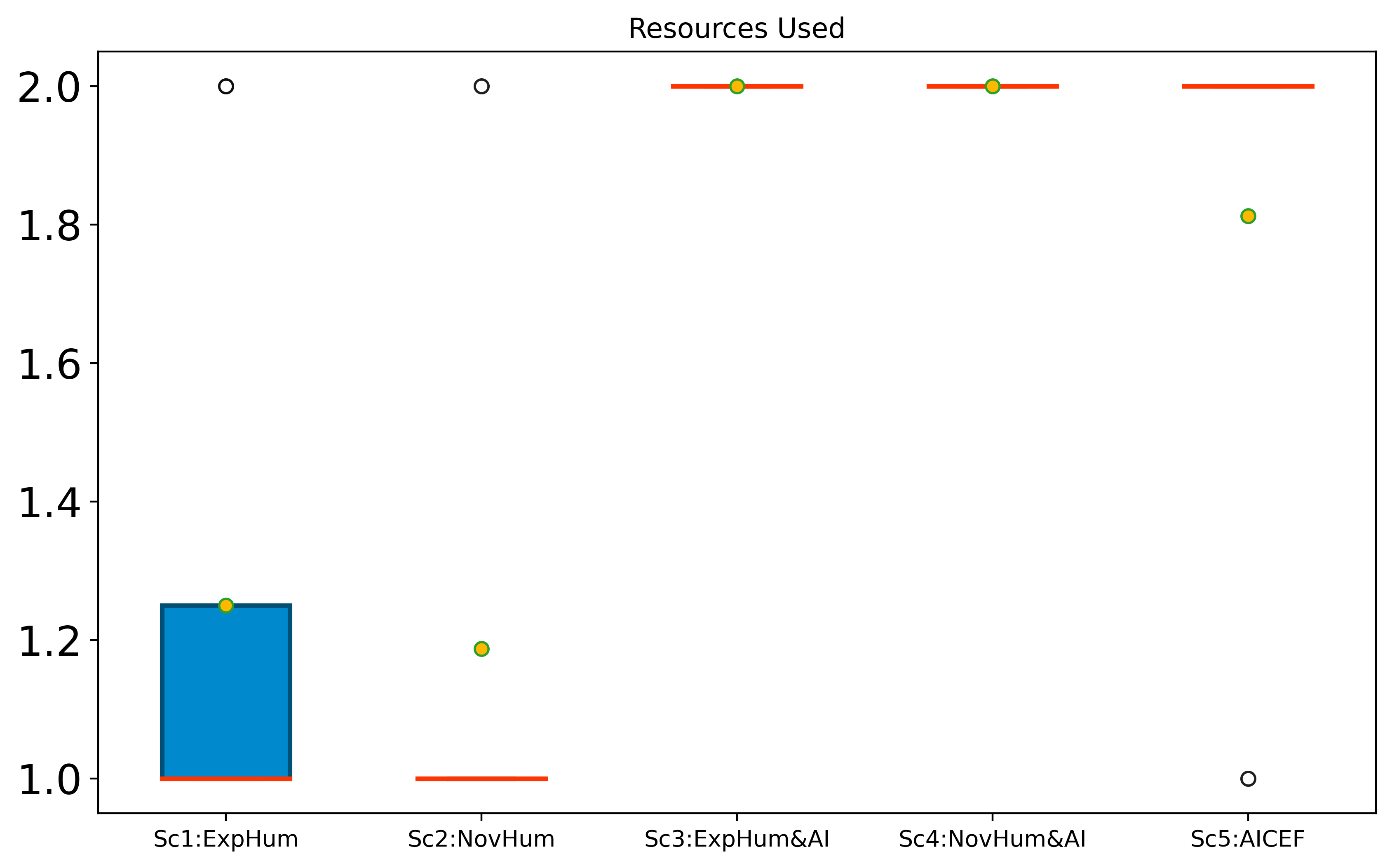}
        \caption{Resources Used}
    \end{subfigure}

    \begin{subfigure}[t]{0.48\textwidth}
        \includegraphics[width=\linewidth]{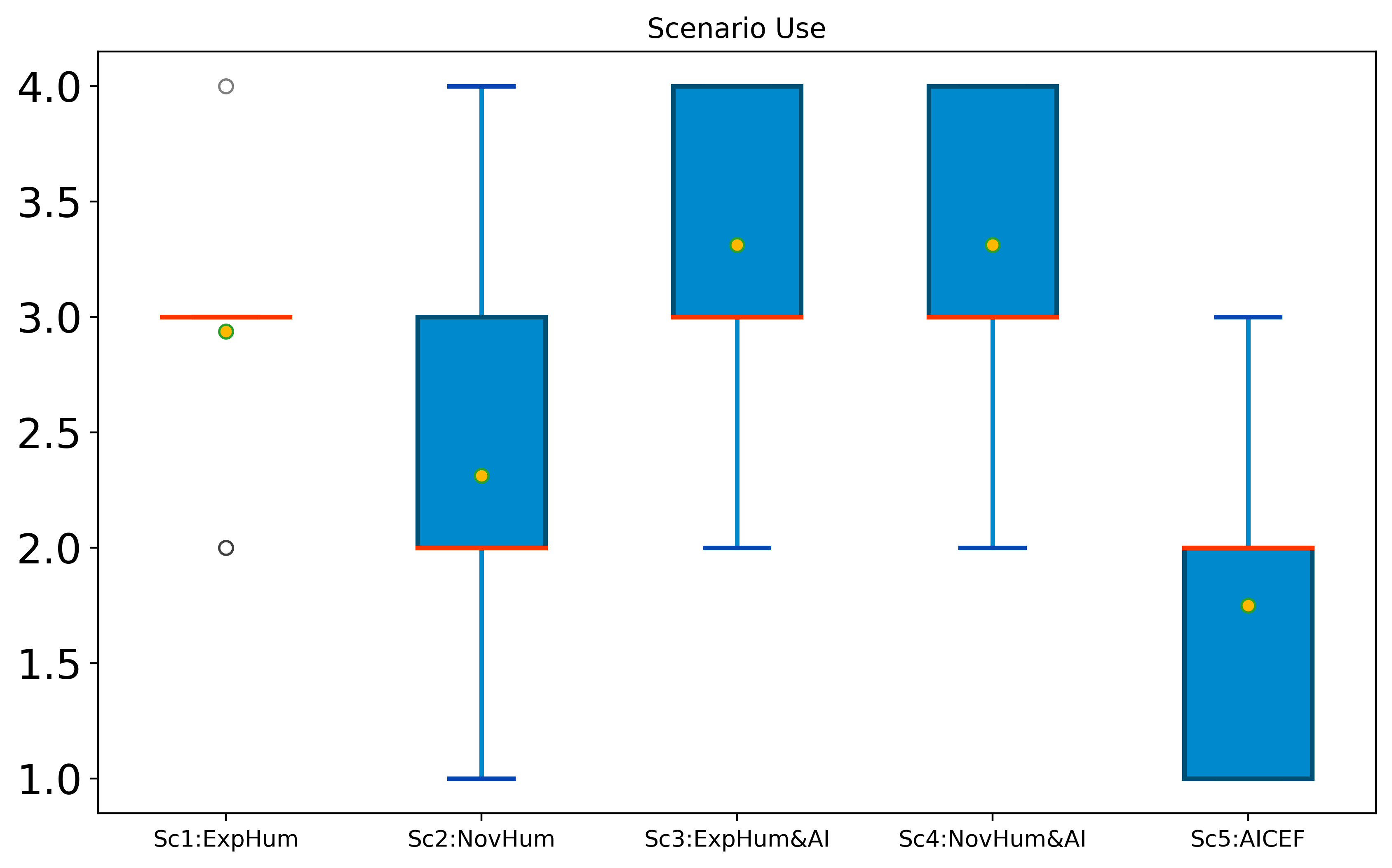}
        \caption{Scenario use.}
    \end{subfigure}
    \begin{subfigure}[t]{0.48\textwidth}
        \includegraphics[width=\linewidth]{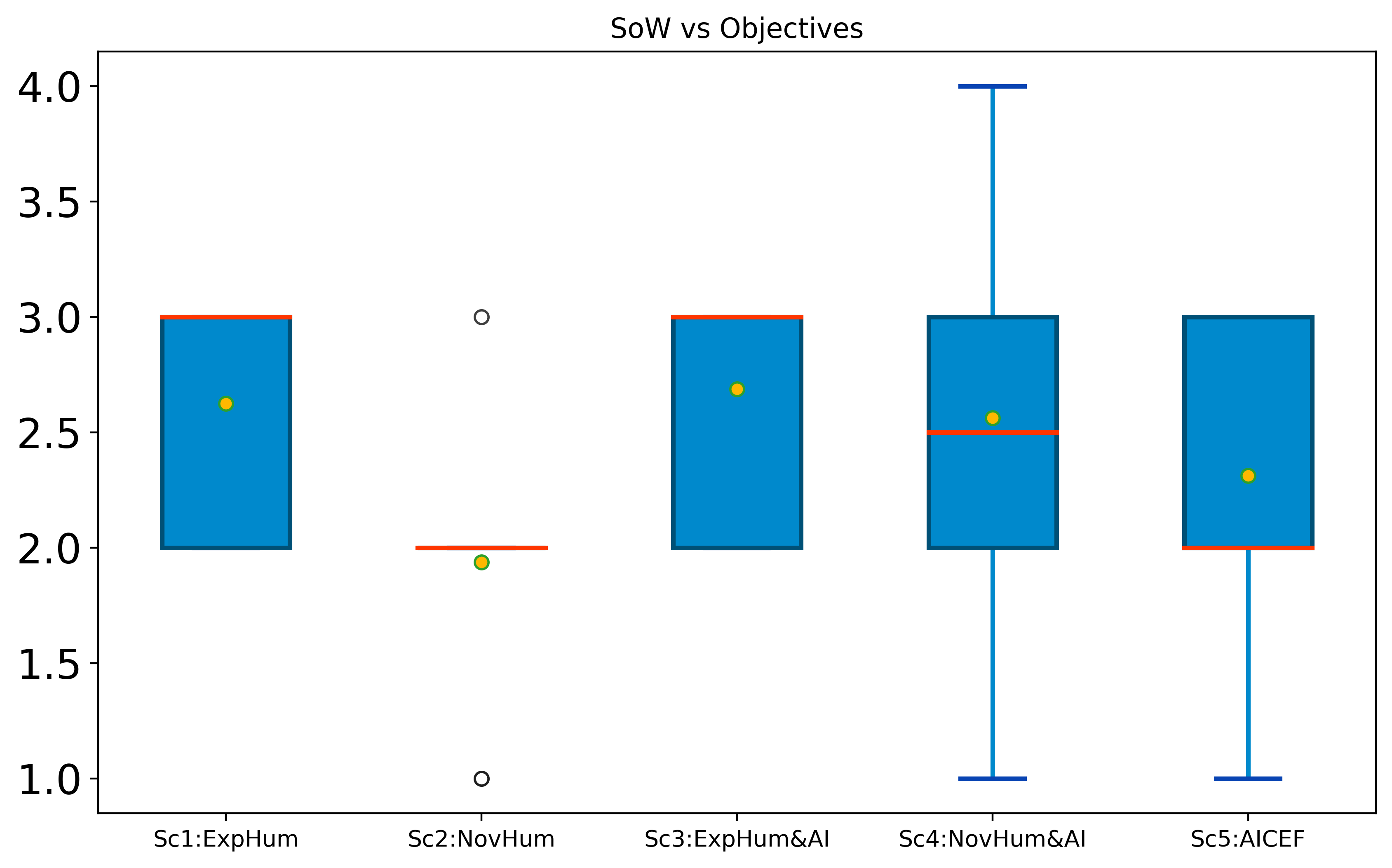}
        \caption{SoW vs Objectives}
    \end{subfigure}

    \begin{subfigure}[t]{0.48\textwidth}
        \includegraphics[width=\linewidth]{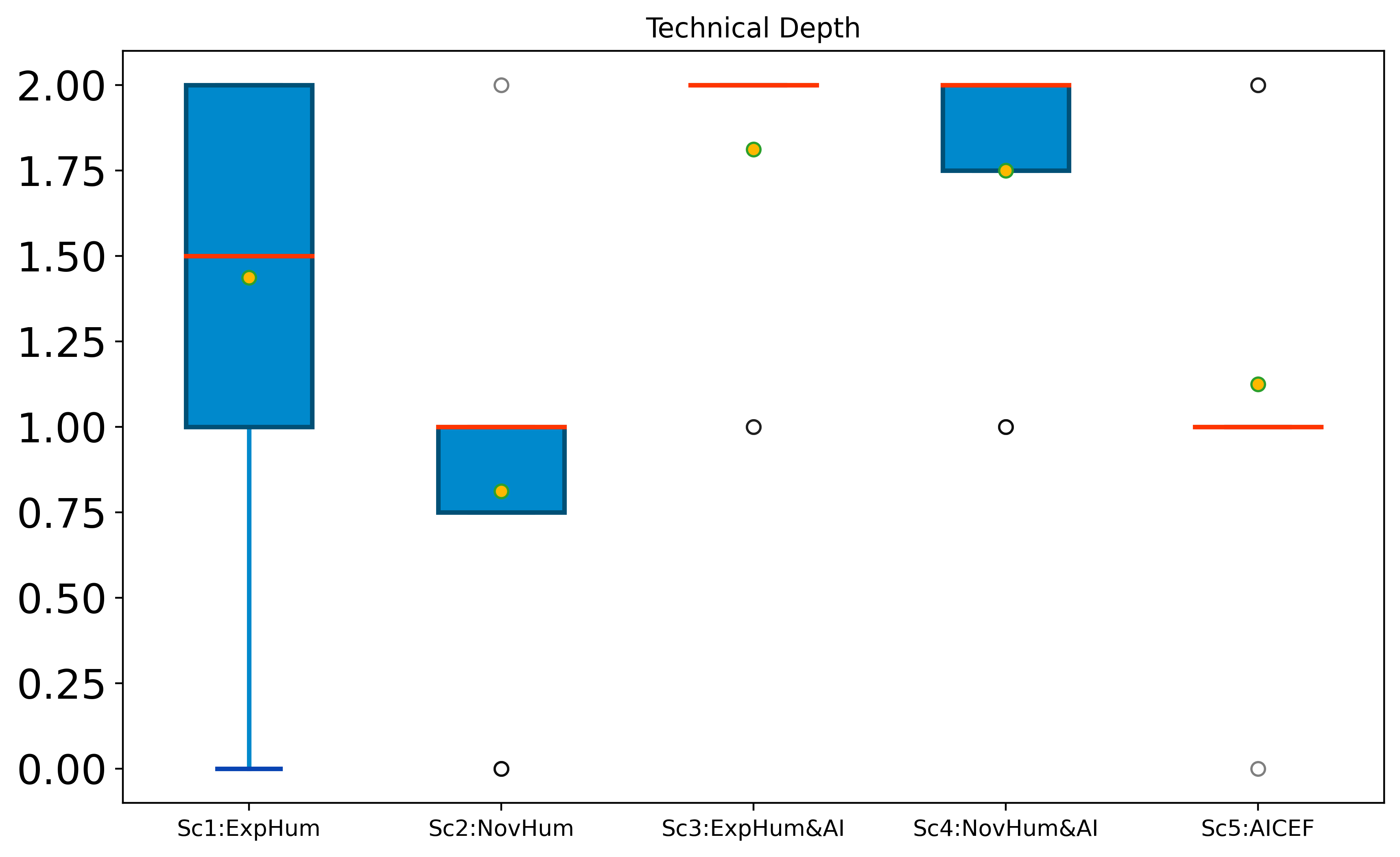}
        \caption{Technical Depth}
    \end{subfigure}
    \begin{subfigure}[t]{0.48\textwidth}
\includegraphics[width=\linewidth]{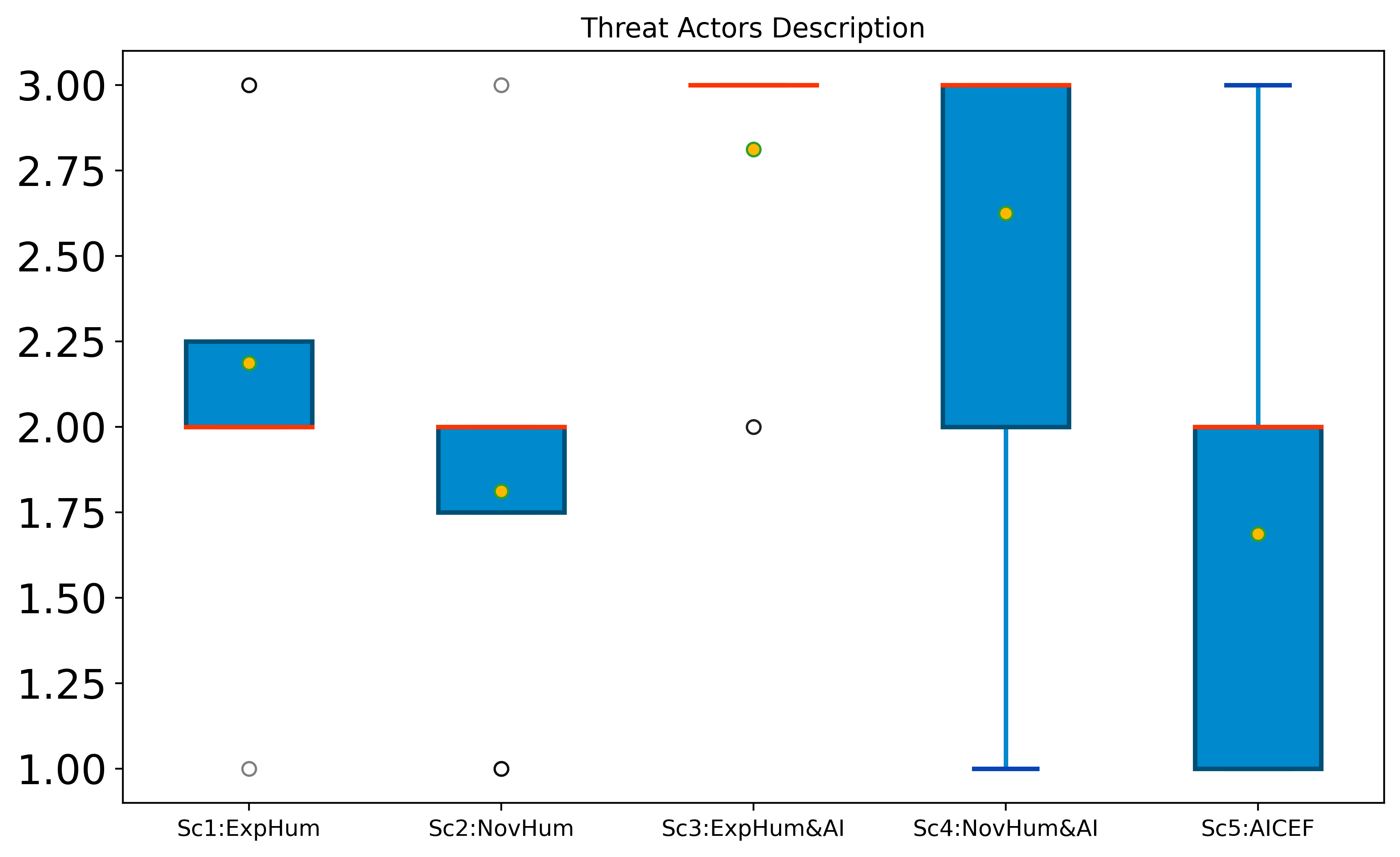}
        \caption{Threat Actors Description}
    \end{subfigure}
    \caption{Range of Scores for the Q1-8 of 16 Evaluators.}
    \label{fig:ranges}
\end{figure*}

Nevertheless, the most impressive finding was that novice planners using AiCEF (Sc4:NovHum\&AI) outperform a seasoned exercise planner (Sc1:ExpHum), as seen in Figure \ref{fig:scenario_eval}, providing a good indication of the capabilities of the proposed framework. Note that the scenario performance developed by the novice planner with the help of AiCEF matches, among others, that of a Seasoned Planner in the question: \textit{Would you use the scenario?}". Even more, evaluators could not distinguish the pure AI-generated content (ExSC5) based on Table \ref{tbl:turing}, categorising the scenario as either hybrid or human-made. Indeed, the results were like those of a novice human planner.

\begin{table}[th]
\centering
\begin{tabular}{|l|r|r|l|}
\hline
\textbf{Scenario} & \textbf{Human} & \textbf{AI \& Human} & \textbf{AI} \\ \hline
Sc1:ExpHum & 1 & 13 & 2 \\ \hline
Sc2:NovHum & 9 & 2 & 5 \\ \hline
Sc3:ExpHum\&AI & 0 & 9 & 7 \\ \hline
Sc4:NovHum\&AI & 1 & 10 & 5 \\ \hline
Sc5:AICEF & 6 & 5 & 5 \\ \hline
\end{tabular}
\caption{Turing Test to evaluate the performance of AI}
\label{tbl:turing}
\end{table}

\begin{figure*}[th]
    \centering
    \includegraphics[width=\textwidth]{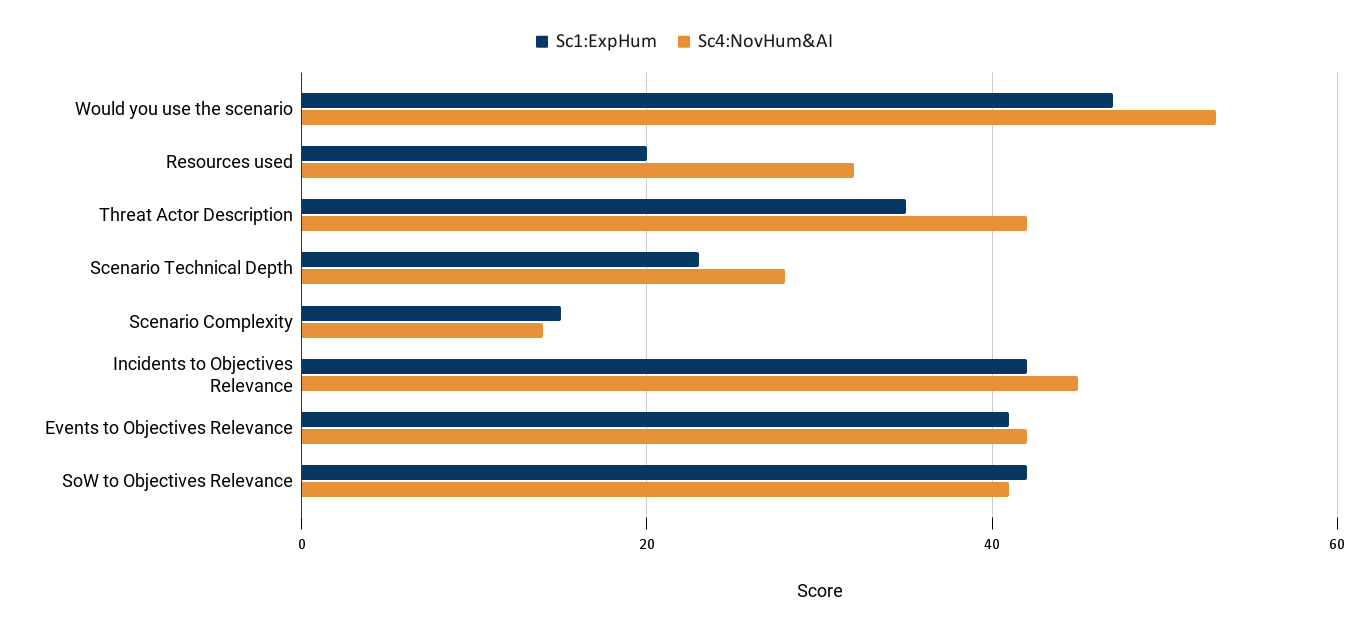}
    \caption{Novice Planner with AiCEF (Sc4:NovHum\&AI) versus Senior Exercise Planner (Sc1:ExpHum)}
    \label{fig:novice_vs_senior}
\end{figure*}

On the question: "\textit{How do you define the scope/objectives of the exercise?"} most evaluators replied with two or more of the following options, with Known incidents \& lessons learnt along with Risk assessment as the most prevalent replies.

\begin{figure*}[th]
    \centering
    \includegraphics[width=\textwidth]{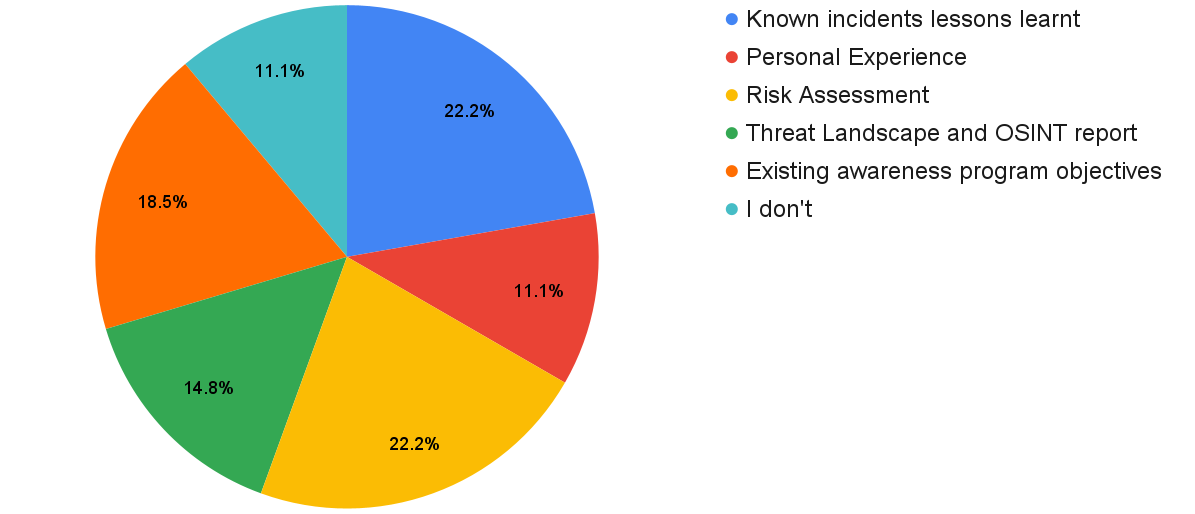}
    \caption{Experts' responses to "How do you define the scope/objectives of the exercise?" question.}
    \label{fig:objectives}
\end{figure*}

On the question: "\textit{How do you define the scenario content?}" most evaluators replied with two or more options, with news and articles being the most important source followed by the known incident option.

\begin{figure*}[th]
    \centering
    \includegraphics[width=.8\textwidth]{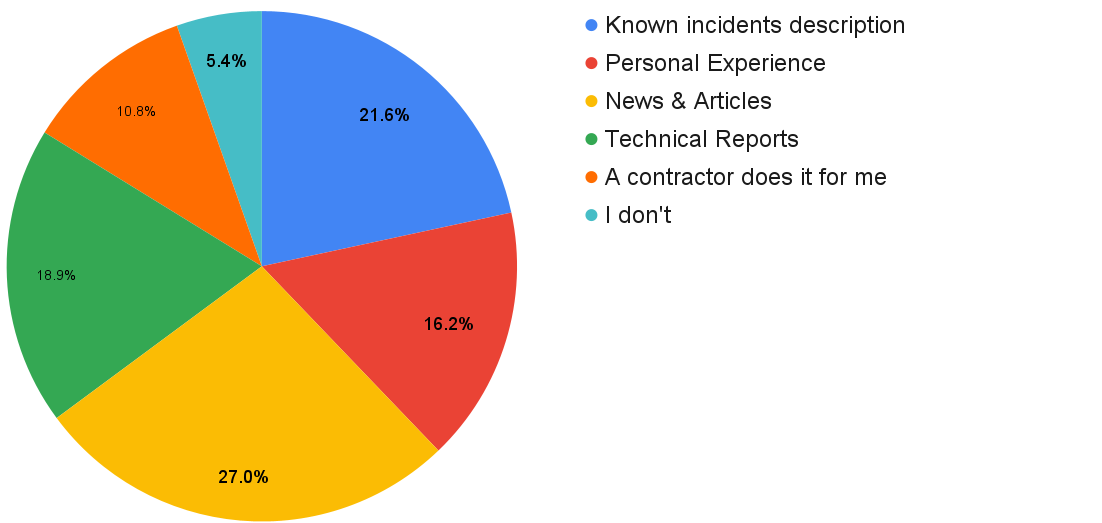}
    \caption{Experts' responses to "How do you define the scenario content?" question.}
    \label{fig:definition}
\end{figure*}

The evaluators replied to the question "\textit{How much time do you invest in the Scenario Content Development?}" with an average of 53 hours. This means that tools which can improve the CSE scenario content development process by reducing time without compromising the quality could be of great use.

Finally, for the question "\textit{What tools did you use to create the scenario or
define the objectives if any?}", the responses varied between Google Search, Cyber Security (News) websites, MS Office, and Internet/Table Top Research.

\section{Conclusions \& Future Work}
The shortage of cybersecurity experts and awareness is a well-known and big worldwide challenge. CSEs can address some of the aspects of this problem; however, the shortage of experts to develop new CSEs coupled with the timeliness and relevance of the developed CSEs requires novel solutions. In this work we try to fill in this gap by facilitating the work of EPs with the use of AI. To this end, we developed a novel AI-powered exercise generation framework called AiCEF which generates structured exercise scenarios that reflect the current or future threat level that an organisation faces, including potential threat actors and TTP's. Moreover, it generates scripted events that could happen in the context of a real attack against a specific organisation belonging in one of the NIS2 critical infrastructure sectors. AiCEF also identifies and describes artefacts that could accompany the exercise scenarios. To this end, AiCEF, uses a new ontology that we built, named CESO, and with which we were able to generate structured exercise scenarios that can be both machine and human-readable.

Our proposed methodology and developed tools can provide tangible qualitative and quantitative added value in CSE development and Cyber Awareness in various ways. For instance, in our experiment, the speed in CSE scenario generation is decreased by 33.33\%  without impacting the  quality. In fact, AiCEF improves the quality of CSE scenario generation for an inexperienced/novice EP, by elevating the generated scenario quality  to the same level of an experienced EP. Finally, the relevance of proposed CSE scenarios is aligned to that of the current threat landscape, as indicated by the evaluation of all the generated scenarios using AiCEF.

While AiCEF might be rather efficient there is room for various improvements. For instance, for operational usage, more sources have to be parsed (ex. threat reports and alerts) to generate more diverse scenarios. While GTP-2 and GTP-3 might create textual output of very good quality, it would be even better if the text synthesizer were based only on Cyber Security related resources so that the generated text is even more relevant and uses, e.g. better technical terms. As indicated in the evaluation, AiCEF could be benefited by further improvements to enhance the threat actor description section. Finally, we plan to enhance AiCEF in order to detect the Cyber Kill Chain phases automatically using NER and create relevant CSE injects for a number of popular categories like Phishing, while also automating the inject description and content generation using AI powered text synthesis.



\section*{Acknowledgements}
This work was supported by the European Commission under the Horizon 2020 Programme (H2020), as part of the project CyberSec4Europe (\url{https://www.cybersec4europe.eu}) (Grant Agreement no. 830929), and Horizon Europe Programme, as part of the project LAZARUS (\url{https://lazarus-he.eu/}) (Grant Agreement no. 101070303) and

\bibliographystyle{plain}
\bibliography{refs}

\end{document}